\documentclass[%
reprint,
superscriptaddress,
 amsmath,amssymb,
 aps,
]{revtex4-2}

\UseRawInputEncoding
\usepackage[dvipdfmx]{graphicx}
\usepackage{dcolumn}
\usepackage{bm}
\usepackage{color}
\usepackage{hyperref}
\usepackage{physics}
\usepackage{orcidlink}

\newcommand{\software}[1]{{\tt #1}}
\newcommand{\mapcube}{\langle\mathcal{M}_{\rm ap}^3\rangle}
\usepackage{afterpage}
\usepackage{lineno}

\begin{document}

\preprint{APS/123-QED}

\title{Dark Energy Survey Year 3 Results: Cosmological constraints from second and third-order shear statistics}

\author{R. C. H. ~Gomes\orcidlink{0000-0002-3800-5662}}
\affiliation{Department of Physics and Astronomy, University of Pennsylvania, Philadelphia, PA 19104, USA}

\author{S. ~Sugiyama\orcidlink{0000-0003-1153-6735}}
\affiliation{Department of Physics and Astronomy, University of Pennsylvania, Philadelphia, PA 19104, USA}

\author{B. ~Jain\orcidlink{0000-0002-8220-3973}}
\affiliation{Department of Physics and Astronomy, University of Pennsylvania, Philadelphia, PA 19104, USA}

\author{M. ~Jarvis}
\affiliation{Department of Physics and Astronomy, University of Pennsylvania, Philadelphia, PA 19104, USA}

\author{D. ~Anbajagane}
\affiliation{Department of Astronomy and Astrophysics, University of Chicago, Chicago, IL 60637, USA}

\author{A. ~Halder}
\affiliation{Institute of Astronomy and Kavli Institute for Cosmology, University of Cambridge, Madingley Road,
Cambridge, CB3 0HA, United Kingdom}
\affiliation{Jesus College, Jesus Lane,
Cambridge, CB5 8BL, United Kingdom}
\affiliation{Max Planck Institute for Extraterrestrial Physics, Giessenbachstra\ss e 1, 85748 Garching,
Germany}
\affiliation{University Observatory, Ludwig-Maximilians-University of Munich, Scheinerstra\ss e 1, 81679 Munich, Germany}

\author{G. A. ~Marques}
\affiliation{Centro Brasileiro de Pesquisas F\'isicas, R. Dr. Xavier Sigaud, 150 - Botafogo, Rio de Janeiro - RJ, 22290-180, Brazil}
\affiliation{Fermi National Accelerator Laboratory, Batavia, IL 60510, USA}
\affiliation{Kavli Institute for Cosmological Physics, University of Chicago, Chicago, IL 60637, USA}

\author{S. ~Pandey}
\affiliation{Columbia Astrophysics Laboratory, Columbia University, 550 West 120th Street, New York, NY 10027, USA}

\author{J. ~Marshall}
\affiliation{George P. and Cynthia Woods Mitchell Institute for Fundamental Physics and Astronomy, and
Department of Physics and Astronomy, Texas A\&M University, College Station, TX 77843, USA}

\author{A.~Alarcon}
\affiliation{Argonne National Laboratory, 9700 South Cass Avenue, Lemont, IL 60439, USA}

\author{A.~Amon}
\affiliation{Department of Astrophysical Sciences, Princeton University, Peyton Hall, Princeton, NJ 08544, USA}

\author{K.~Bechtol}
\affiliation{Physics Department, 2320 Chamberlin Hall, University of Wisconsin-Madison, 1150 University Avenue Madison, WI 53706-1390}

\author{M.~Becker}
\affiliation{Argonne National Laboratory, 9700 South Cass Avenue, Lemont, IL 60439, USA}

\author{G.~Bernstein}
\affiliation{Department of Physics and Astronomy, University of Pennsylvania, Philadelphia, PA 19104, USA}

\author{A.~Campos}
\affiliation{Department of Physics, Carnegie Mellon University, Pittsburgh, PA 15312, USA}

\author{R.~Cawthon}
\affiliation{Physics Department, William Jewell College, Liberty, MO 64068, USA}

\author{C.~Chang}
\affiliation{Kavli Institute for Cosmological Physics, University of Chicago, Chicago, IL 60637, USA}
\affiliation{Department of Astronomy and Astrophysics, University of Chicago, Chicago, IL 60637, USA}

\author{R.~Chen}
\affiliation{Department of Physics, Duke University, Durham, NC 27708, USA}

\author{A.~Choi}
\affiliation{NASA Goddard Space Flight Center, 8800 Greenbelt Road, Greenbelt, Maryland 20771, USA}

\author{J.~Cordero}
\affiliation{Jodrell Bank Center for Astrophysics, School of Physics and Astronomy, University of Manchester, Oxford Road, Manchester M13 9PL, United Kingdom}

\author{C.~Davis}
\affiliation{Kavli Institute for Particle Astrophysics \& Cosmology, P. O. Box 2450, Stanford University, Stanford, CA 94305, USA}

\author{J.~Derose}
\affiliation{Lawrence Berkeley National Laboratory, 1 Cyclotron Road, Berkeley, California 94720, USA}

\author{S.~Dodelson}
\affiliation{McWilliams Center for Cosmology and Astrophysics, Department of Physics, Carnegie Mellon University, Pittsburgh, PA 15213, USA}
\affiliation{Department of Astronomy and Astrophysics, University of Chicago, Chicago, IL 60637, USA}
\affiliation{Fermi National Accelerator Laboratory, P. O. Box 500, Batavia, IL 60510, USA}
\affiliation{Kavli Institute for Cosmological Physics, University of Chicago, Chicago, IL 60637, USA}

\author{C.~Doux}
\affiliation{Universit\'e Grenoble Alpes, CNRS, LPSC-IN2P3, 38000 Grenoble, France}

\author{K.~Eckert}
\affiliation{Department of Physics and Astronomy, University of Pennsylvania, Philadelphia, PA 19104, USA}

\author{F.~Elsner}
\affiliation{Department of Physics \& Astronomy, University College London, Gower Street, London, WC1E 6BT, UK}

\author{J.~Elvin-Poole}
\affiliation{Department of Physics and Astronomy, University of Waterloo, 200 University Ave W, Waterloo, ON N2L 3G1, Canada}
\affiliation{Waterloo Centre for Astrophysics, University of Waterloo, 200 University Ave W, Waterloo, ON N2L 3G1, Canada}

\author{S.~Everett}
\affiliation{California Institute of Technology, 1200 East California Boulevard, Pasadena, CA 91125, USA}

\author{A.~Fert\'e}
\affiliation{SLAC National Accelerator Laboratory, Menlo Park, CA, USA}
\affiliation{Kavli Institute for Particle Astrophysics and Cosmology, Stanford University, Stanford, CA, USA}

\author{M.~Gatti}
\affiliation{Kavli Institute for Cosmological Physics, University of Chicago, Chicago, IL 60637, USA}

\author{G.~Giannini}
\affiliation{Institut de F\'ısica d’Altes Energies (IFAE), The Barcelona Institute of Science and Technology, Campus UAB, 08193 Bellaterra (Barcelona) Spain}
\affiliation{Kavli Institute for Cosmological Physics, University of Chicago, 5640 South Ellis Avenue, Chicago, IL 60637, USA}

\author{D.~Gruen}
\affiliation{Universit\"ats-Sternwarte, Fakult\"at f\"ur Physik, Ludwig-Maximilians-Universit\"at M\"unchen, Scheinerstra\ss e 1, 81679 M\"unchen, Germany}
\affiliation{Excellence Cluster ORIGINS, Boltzmannstra\ss e 2, 85748 Garching, Germany}

\author{I.~Harrison}
\affiliation{School of Physics and Astronomy, Cardiff University, The Parade, Cardiff, Wales CF24 3AA, UK}

\author{K.~Herner}
\affiliation{Fermi National Accelerator Laboratory, P. O. Box 500, Batavia, IL 60510, USA}

\author{E.~M.~Huff}
\affiliation{Jet Propulsion Laboratory, California Institute of Technology, 4800 Oak Grove Dr., Pasadena, CA 91109, USA}

\author{D.~Huterer}
\affiliation{Department of Physics, University of Michigan, 450 Church Street, Ann Arbor, MI 48109, USA}
\affiliation{University of Michigan, 500 S. State Street, Ann Arbor, MI 48109, USA}

\author{N.~Kuropatkin}
\affiliation{Fermi National Accelerator Laboratory, P. O. Box 500, Batavia, IL 60510, USA}

\author{P.~F.~Leget}
\affiliation{LPNHE, CNRS/IN2P3, Sorbonne Universite, Laboratoire de Physique Nucl\'eaire et de Hautes Energies, F-75005, Paris, France}

\author{N.~Maccrann}
\affiliation{DAMTP, Centre for Mathematical Sciences, University of Cambridge, Wilberforce Road, Cambridge CB3 0WA, UK}
\affiliation{Kavli Institute for Cosmology Cambridge, Madingley Road, Cambridge CB3 0HA, UK}

\author{J.~Mccullough}
\affiliation{Department of Astrophysical Sciences, Princeton University, Peyton Hall, Princeton, NJ 08544, USA}
\affiliation{Kavli Institute for Particle Astrophysics \& Cosmology, P.O. Box 2450, Stanford University, Stanford, CA 94305, USA}
\affiliation{SLAC National Accelerator Laboratory, Menlo Park, CA 94025, USA}

\author{J.~Muir}
\affiliation{Kavli Institute for Particle Astrophysics \& Cosmology, P. O. Box 2450, Stanford University, Stanford, CA 94305, USA}

\author{J.~Myles}
\affiliation{Department of Astrophysical Sciences, Princeton University, Peyton Hall, Princeton, NJ 08544, USA}

\author{A.~Navarro~Alsina}
\affiliation{Instituto de F\'isica Gleb Wataghin, Universidade Estadual de Campinas, 13083-859, Campinas, SP, Brazil}
\affiliation{Laborat\'orio Interinstitucional de e-Astronomia, Rua Gal. Jos\'e Cristino 77, Rio de Janeiro, RJ - 20921-400, Brazil}

\author{J.~Prat}
\affiliation{Nordita, Stockholm University and KTH Royal Institute of Technology, Hannes Alfv\'ens v\"ag 12, SE-10691 Stockholm, Sweden}

\author{M.~Raveri}
\affiliation{Department of Physics and INFN, University of Genova, Genova, Italy}

\author{R.~P.~Rollins}
\affiliation{Jodrell Bank Centre for Astrophysics, School of Physics and Astronomy, University of Manchester, Oxford Road, Manchester, M13 9PL, United Kingdom}

\author{A.~Roodman}
\affiliation{Kavli Institute for Particle Astrophysics \& Cosmology, P.O. Box 2450, Stanford University, Stanford, CA 94305, USA}
\affiliation{SLAC National Accelerator Laboratory, Menlo Park, CA 94025, USA}

\author{A.~J.~Ross}
\affiliation{Center for Cosmology and Astro-Particle Physics, The Ohio State University, Columbus, OH 43210, USA}

\author{E.~S.~Rykoff}
\affiliation{Kavli Institute for Particle Astrophysics \& Cosmology, P. O. Box 2450, Stanford University, Stanford, CA 94305, USA}
\affiliation{SLAC National Accelerator Laboratory, Menlo Park, CA 94025, USA}

\author{C.~S\'anchez}
\affiliation{Departament de F\'isica, Universitat Aut\`onoma de Barcelona (UAB), 08193 Bellaterra (Barcelona), Spain}
\affiliation{Institut de F\'isica d’Altes Energies (IFAE), The Barcelona Institute of Science and Technology, Campus UAB, 08193 Bellaterra (Barcelona), Spain}

\author{L.~F.~Secco}
\affiliation{Kavli Institute for Cosmological Physics, University of Chicago, Chicago, IL 60637, USA}

\author{E.~Sheldon}
\affiliation{Brookhaven National Laboratory, Bldg 510, Upton, NY 11973, USA}

\author{T.~Shin}
\affiliation{Department of Physics, Carnegie Mellon University, Pittsburgh, PA 15213, USA}

\author{M.~Troxel}
\affiliation{Department of Physics, Duke University, Durham, NC 27708, USA}

\author{I.~Tutusaus}
\affiliation{Institut de Recherche en Astrophysique et Plan\'etologie (IRAP), Universit\'e de Toulouse, CNRS, UPS, CNES, 14 Av. Edouard Belin, 31400 Toulouse, France}

\author{T.~N.~Varga}
\affiliation{Excellence Cluster Origins, Boltzmannstr. 2, 85748 Garching, Germany}
\affiliation{Max Planck Institute for Extraterrestrial Physics, Giessenbachstrasse, 85748 Garching, Germany}
\affiliation{Universit\"ats-Sternwarte, Fakult\"at f\"ur Physik, LudwigMaximilians Universit\"at M\"unchen, Scheinerstr. 1, 81679 M\"unchen, Germany}

\author{B.~Yanny}
\affiliation{Fermi National Accelerator Laboratory, PO Box 500, Batavia, IL, 60510, USA}

\author{B.~Yin}
\affiliation{Department of Physics, Duke University Durham, NC 27708, USA}

\author{Y.~Zhang}
\affiliation{NSF National Optical-Infrared Astronomy Research Laboratory, 950 N Cherry Avenue, Tucson, AZ 85719}

\author{J.~Zuntz}
\affiliation{Institute for Astronomy, University of Edinburgh, Royal Observatory, Blackford Hill, Edinburgh, EH9 3HJ, UK}

\author{M. ~Aguena}
\affiliation{INAF-Osservatorio Astronomico di Trieste, via G. B. Tiepolo 11, I-34143 Trieste, Italy}
\affiliation{Laborat\'orio Interinstitucional de e-Astronomia - LIneA, Av. Pastor Martin Luther King Jr, 126 Del Castilho, Nova Am\'erica Offices,
Torre 3000/sala 817 CEP: 20765-000, Brazil}

\author{F. ~Andrade-Oliveira}
\affiliation{Physik-Institut, University of Z\"urich, Winterthurerstrasse 190, CH-8057 Z\"urich, Switzerland}

\author{D. ~Bacon}
\affiliation{Institute of Cosmology and Gravitation, University of Portsmouth,
Portsmouth, PO1 3FX, UK}

\author{J. ~Blazek}
\affiliation{Department of Physics, Northeastern University, Boston, MA 02115, USA}

\author{S. ~Bocquet}
\affiliation{University Observatory, LMU Faculty of Physics, Scheinerstr. 1, 81679 Munich, Germany}

\author{D. ~Brooks}
\affiliation{Department of Physics \& Astronomy, University College London, Gower Street, London, WC1E 6BT, UK}

\author{A. ~Carnero Rosell}
\affiliation{ Instituto de Astrofisica de Canarias, E-38205 La Laguna, Tenerife, Spain}
\affiliation{Laborat\'orio Interinstitucional de e-Astronomia - LIneA, Av. Pastor Martin Luther King Jr, 126 Del Castilho, Nova Am\'erica Offices,
Torre 3000/sala 817 CEP: 20765-000, Brazil}
\affiliation{Universidad de La Laguna, Dpto. Astrof\'isica, E-38206 La Laguna, Tenerife, Spain}

\author{J. ~Carretero}
\affiliation{Institut de F\'isica d’Altes Energies (IFAE), The Barcelona Institute of Science and Technology, Campus UAB, 08193 Bellaterra
(Barcelona) Spain}

\author{M. ~Costanzi}
\affiliation{Astronomy Unit, Department of Physics, University of Trieste, via Tiepolo 11, I-34131 Trieste, Italy}
\affiliation{INAF-Osservatorio Astronomico di Trieste, via G. B. Tiepolo 11, I-34143 Trieste, Italy}
\affiliation{Institute for Fundamental Physics of the Universe, Via Beirut 2, 34014 Trieste, Italy}

\author{L. ~da Costa}
\affiliation{Laborat\'orio Interinstitucional de e-Astronomia - LIneA, Av. Pastor Martin Luther King Jr, 126 Del Castilho, Nova Am\'erica Offices,
Torre 3000/sala 817 CEP: 20765-000, Brazil}

\author{M. E. ~da Silva Pereira}
\affiliation{Hamburger Sternwarte, Universit\"at Hamburg, Gojenbergsweg 112, 21029 Hamburg, Germany}

\author{T. M. ~Davis}
\affiliation{School of Mathematics and Physics, University of Queensland, Brisbane, QLD 4072, Australia}

\author{J. ~De Vicente}
\affiliation{Centro de Investigaciones Energ\'eticas, Medioambientales y Tecnol\'ogicas (CIEMAT), Madrid, Spain}

\author{H. T. ~Diehl}
\affiliation{Fermi National Accelerator Laboratory, P.O. Box 500, Batavia, IL 60510, USA}

\author{B. ~Flaugher}
\affiliation{Fermi National Accelerator Laboratory, P.O. Box 500, Batavia, IL 60510, USA}

\author{J. ~Frieman}
\affiliation{Department of Astronomy and Astrophysics, University of Chicago, Chicago, IL 60637, USA}
\affiliation{Fermi National Accelerator Laboratory, P.O. Box 500, Batavia, IL 60510, USA}
\affiliation{Kavli Institute for Cosmological Physics, University of Chicago, Chicago, IL 60637, USA}

\author{G. ~Gutierrez}
\affiliation{Fermi National Accelerator Laboratory, P.O. Box 500, Batavia, IL 60510, USA}

\author{S. R. ~Hinton}
\affiliation{School of Mathematics and Physics, University of Queensland, Brisbane, QLD 4072, Australia}

\author{D. L. ~Hollowood}
\affiliation{Santa Cruz Institute for Particle Physics, Santa Cruz, CA 95064, USA}

\author{K. ~Honscheid}
\affiliation{Center for Cosmology and AstroParticle Physics, The Ohio State University, 191 West Woodruff Avenue, Columbus, OH 43210, USA}
\affiliation{Department of Physics, The Ohio State University, 191 West Woodruff Avenue, Columbus, OH 43210, USA}
\affiliation{The Ohio State University, Columbus, 43210 OH, USA}

\author{D. J. ~James}
\affiliation{Center for Astrophysics | Harvard \& Smithsonian, 60 Garden Street, Cambridge, MA 02138, USA}

\author{N. ~Jeffrey}
\affiliation{Department of Physics \& Astronomy, University College London, Gower Street, London, WC1E
6BT, UK}

\author{S. ~Lee}
\affiliation{Jet Propulsion Laboratory, California Institute of Technology, 4800 Oak Grove Dr., Pasadena, CA
91109, USA}

\author{J. ~Mena-Fern\'andez}
\affiliation{LPSC Grenoble - 53, Avenue des Martyrs 38026 Grenoble, France}

\author{R. ~Miquel}
\affiliation{Instituci\'o Catalana de Recerca i Estudis Avan\c{c}ats, E-08010 Barcelona, Spain}
\affiliation{Institut de F\'isica d’Altes Energies (IFAE), The Barcelona Institute of Science and Technology,
Campus UAB, 08193 Bellaterra (Barcelona) Spain}

\author{R. L. C. ~Ogando}
\affiliation{Centro de Tecnologia da Informa\c{c}\~ao Renato Archer, Campinas, SP, Brazil - 13069-901}
\affiliation{Observat\'orio Nacional, Rio de Janeiro, RJ, Brazil - 20921-400}

\author{A. A. ~Plazas Malag\'on}
\affiliation{Kavli Institute for Particle Astrophysics \& Cosmology, P. O. Box 2450, Stanford University, Stanford, CA 94305, USA}
\affiliation{SLAC National Accelerator Laboratory, Menlo Park, CA 94025, USA}

\author{A. ~Porredon}
\affiliation{Centro de Investigaciones Energ\'eticas, Medioambientales y Tecnol\'ogicas (CIEMAT), Madrid, Spain}
\affiliation{Ruhr University Bochum, Faculty of Physics and Astronomy, Astronomical Institute, German Centre for Cosmological Lensing, 44780 Bochum, Germany}

\author{E. ~Sanchez}
\affiliation{Centro de Investigaciones Energ\'eticas, Medioambientales y Tecnol\'ogicas (CIEMAT), Madrid, Spain}

\author{D. ~Sanchez Cid}
\affiliation{Centro de Investigaciones Energ\'eticas, Medioambientales y Tecnol\'ogicas (CIEMAT), Madrid, Spain}

\author{S. ~Samuroff}
\affiliation{Department of Physics, Northeastern University, Boston, MA 02115, USA}

\author{M. ~Smith}
\affiliation{School of Physics and Astronomy, University of Southampton, Southampton, SO17 1BJ, UK}

\author{E. ~Suchyta}
\affiliation{Computer Science and Mathematics Division, Oak Ridge National Laboratory, Oak Ridge, TN 37831}

\author{M. E. C. ~Swanson}
\affiliation{Center for Astrophysical Surveys, National Center for Supercomputing Applications, 1205 West Clark St., Urbana, IL 61801, USA}

\author{D. ~Thomas}
\affiliation{Institute of Cosmology and Gravitation, University of Portsmouth,
Portsmouth, PO1 3FX, UK}

\author{V.~Vikram}
\affiliation{Department of Physics, Central University of Kerala, 93VR+RWF, CUK Rd, Kerala 671316, India}

\author{J.~Weller}
\affiliation{Max Planck Institute for Extraterrestrial Physics, Giessenbachstrasse, 85748 Garching, Germany}
\affiliation{Universit\"ats-Sternwarte, Fakult\"at f\"ur Physik, Ludwig-Maximilians Universit\"at M\"unchen, Scheinerstr. 1, 81679 M\"unchen, Germany}

\author{M. ~Yamamoto}
\affiliation{Department of Astrophysical Sciences, Princeton University, Peyton Hall, Princeton, NJ 08544, USA}

\collaboration{DES Collaboration}

\date{\today}

\begin{abstract}
We present a cosmological analysis of the third-order aperture mass statistic using Dark Energy Survey Year 3 (DES Y3) data. We perform a complete tomographic measurement of the three-point correlation function of the Y3 weak lensing shape catalog with the four fiducial  source redshift bins. Building upon our companion methodology paper, we apply a pipeline that combines the two-point function $\xi_{\pm}$ with the mass aperture skewness statistic $\mapcube$, which is an efficient compression of the full shear three-point function. We use a suite of simulated shear maps to obtain a joint covariance matrix. By jointly analyzing $\xi_\pm$ and $\mapcube$ measured from DES Y3 data with a $\Lambda$CDM model,
we find $S_8=0.780\pm0.015$ and $\Omega_{\rm m}=0.266^{+0.039}_{-0.040}$, yielding 111\% of figure-of-merit improvement in $\Omega_m$-$S_8$ plane relative to $\xi_{\pm}$ alone, consistent with expectations from simulated likelihood analyses. With a $w$CDM model, we find $S_8=0.749^{+0.027}_{-0.026}$ and $w_0=-1.39\pm 0.31$, which gives an improvement of $22\%$ on the joint $S_8$-$w_0$ constraint. Our results are consistent with $w_0=-1$. Our new constraints are compared to CMB data from the Planck satellite, and we find that with the inclusion of $\mapcube$  the existing tension between the data sets is at the level of $2.3\sigma$. We show that the third-order statistic enables us to self-calibrate the mean photometric redshift uncertainty parameter of the highest redshift bin with little degradation in the figure of merit. Our results demonstrate the constraining power of higher-order lensing statistics and establish $\mapcube$ as a practical observable for joint analyses in current and future surveys.
\end{abstract}

\maketitle

\section{Introduction}\label{sec:introduction}

The large-scale structure of the Universe has been extensively studied over the past decades by a variety of cosmological surveys. In this context, the analysis of cosmic shear has proven to be a powerful tool in placing competitive constraints on cosmological theory models \cite{Secco.Weller.2022,Amon.Weller.2021,Asgari.Valentijn.2020,Dalal.Wang.2023,Li.Wang.2023}. The traditional way to perform such analyses is by measuring second-order summary statistics, such as the two-point correlation function (2PCF), on survey catalogs, and subsequently testing the model predictions for the statistic being used. However, second-order statistics are not able to capture the rich non-Gaussian information imprinted by the non-linear growth of cosmic structure. 

The cosmic shear three-point correlation function (3PCF) was studied by \citet{Schneider.Lombardi.2002} and proposed as a complement to the two-point function that allows the probing of non-Gaussian features of the field. Other higher-order statistics (HOS) have since been proposed \cite{Petri_2013,Cheng.Yuan-Sen,Allys_2020,barthelemy2024makingleapimodelling, Zurcher.Fluri.2022, Giblin_2023}  and applied to mock and survey data, obtaining different levels of improvement over their corresponding second-order constraints \cite{Marques_2024,Heydenreich_2021_homology,anbajagane20233rdmomentpracticalstudy, Gong_2023, Jeffrey.Whiteway.Gatti.2024, Prat.Gatti.Doux.2025}. Most of the available HOS are constructed from simulated maps, not being analytically built from theoretical models.

High signal-to-noise (S/N) measurements of the 3PCF on Dark Energy Survey Year 3 (DES Y3) data have been performed by \citet{Secco.Weller.2022}, showing the potential of this data set to undergo a joint 2PCF and 3PCF analysis. Such an analysis has the advantage of being directly modeled from theory, and also allows us a more direct study of the contributions of different scales. This stands in contrast to the approach of \citet{Gatti.Collaboration.2022}, which requires mass map reconstruction from the Y3 shear catalog in order to extract the third moment of the convergence field. Our previous work has established a fast algorithm to compute the model prediction of the three-point correlation function \cite{sugiyama2024fastmodelingshearthreepoint}, obtaining an improvement on runtime of six orders of magnitude relative to brute-force integration \cite[see also][for a similar idea for the fast modeling of three-point correlation function]{Arvizu:2024rlt,Samario-Nava:2025auc}. We make use of these advances and build a pipeline for a full cosmological analysis of the shear two-point functions $\xi_{\pm}$ and of the $\mapcube$ statistic, which is shown to be an efficient compression of the full 3PCF \cite{Heydenreich.Schneider.2022}, on DES Y3 data.

In a companion paper, which we will refer to as Paper~I \cite{gomes2025cosmologysecondthirdordershear}, we have 
developed the methodology, built the joint covariance matrix, and performed modeling of systematics along with thorough simulated likelihood analyses. We have found our methodology to be competitive with other higher-order analyses, yielding a factor-of-two improvement on the joint $\Omega_m$-$S_8$ constraint. This level of improvement is similar to that found by \citet{Gatti.Wiseman.2023} through simulation-based inference combining scattering transforms, wavelet phase harmonics, and third moments of the convergence map. It is also comparable to the $\mapcube$ analysis results from the KiDS collaboration \cite{Burger.Martinet.2023}. In this paper, we apply our methodology to DES Y3 data and obtain improved constraints on $\Omega_m$ and $S_8$, maintaining our predicted factor-of-two improvement on the joint constraint.

The structure of this paper is as follows. In Section II, we present our data and covariance matrix. In Section III, we review the theoretical modeling of the weak lensing summary statistics. In Section IV, we present our measurements of the three-point correlation function and of the mass aperture statistic in the Y3 shape catalog. In Section V, we briefly review our parameter inference methodology and present our blinding procedure. In Section VI, we present and discuss our cosmological constraints, with concluding remarks in Section VII.

\section{Data}\label{sec:data}
\subsection{DES Y3 shape catalog}\label{sec:desy3data}

For this study, we use the DES Y3 weak lensing shape catalog described by \citet{Gatti.Wilkinson.2020} \cite{DES_data}. This catalog was produced from the first three years of data collected through the DECam, which operated on the Blanco 4m telescope, located on the Cerro Tololo Inter-American Observatory. The shear was measured in the data through the METACALIBRATION pipeline, resulting in a catalog with 100,204,026
galaxies. The effective area of the sky covered through these observations is 4143 $\deg^2$, and the effective number density is 5.59 $\text{gal/arcmin}^2$.

The catalog is split into four tomographic redshift bins, with each bin being characterized by a redshift distribution function $n(z)$. The distribution functions for all redshift bins are presented by \citet{Secco.To.2021}. These functions are obtained through the self-organizing map $p(z)$ (SOMPZ) method, and combined with estimates from clustering redshifts and shear ratios, as described by \citet{Myles_2021}. Each of the four bins has a similar number density of galaxies.

\subsection{CosmoGrid simulations for covariance measurement}\label{sec:cosmogrid}

We use Y3 weak lensing data to obtain cosmological constraints from the shear two-point and three-point correlation functions. As described in Section \ref{sec:model}, we perform an analysis with a joint 2PCF and mass aperture data vector.

In order to obtain adequate posteriors for our parameter inference process, we need to compute a joint covariance of our full data vector, which concatenates the two-point and three-point information. This covariance can be estimated either analytically, or from a large set of simulations, or through jackknife sampling of the real data. For this analysis, we use the same covariance matrix described in Paper~I, which includes the second-order shear covariance ($\xi_\pm$), the third-order mass aperture covariance ($\mapcube$), and their cross-covariance. The full matrix is estimated using 796 mock realizations generated from the \software{CosmoGridV1} fiducial cosmology simulations \citep{Kacprzak.Stadel.2022}. Each realization consists of a DES-Y3-like cosmic shear map, cutting a DES-Y3 footprint from a full-sky simulated shear field and adding shape noise derived from the DES-Y3 shape catalog \citep{Gatti.Wilkinson.2020}.

Measurements of $\xi_\pm$ and $\mapcube$ are performed using \software{TreeCorr}, with the angular binning and filters described in Paper~I. The sample covariance is then computed as:
\begin{align}
    \bm{C} = \frac{1}{N_{\rm real}-1}\sum_{r=1}^{N_{\rm real}} \left[\bm{d}^{r} - \bar{\bm{d}}\right] \left[\bm{d}^{r} - \bar{\bm{d}}\right]^{\rm T},
\end{align}
where $\bm{d}^{r}$ is the joint data vector of the $r$-th realization, $\bar{\bm{d}}$ is the mean data vector, and $N_{\rm real}=796$.

To ensure the robustness of the covariance model, we apply scale cuts to the three-point correlation function prior to compressing it into $\mapcube$. We remove contributions from triangles with side lengths below $\theta \approx 8'$. This addresses inaccuracies caused by the resolution of the simulated maps. The impact of this cut and other validation checks, including comparisons with analytic models for $\xi_\pm$ covariance \cite{Amon.Weller.2021}, along with tests of baryonic feedback, are detailed in Paper~I. We verify the convergence of our covariance matrix by performing a simulated analysis from a subsample of our total simulations (700 out of 796) and comparing the final constraints with our fiducial simulated analysis. The number of simulations is also sufficiently larger than our data vector length, which, after compression, equals 96.

We confirm that the cross-covariance between $\xi_\pm$ and $\mapcube$ is negligible, with no significant structure in the matrix. The correlation coefficient has a maximum value of 0.13, consistent with the low overlap in information content between the two statistics.


\section{Weak lensing theory}\label{sec:model}
\subsection{Convergence and Shear}\label{sec:}
Weak gravitational lensing is a powerful probe of the universe's matter distribution, sensitive to both visible and dark matter. The convergence field, $\kappa(\bm{\theta})$ quantifies isotropic magnification of background galaxy shapes and is given by the line-of-sight integration of the matter density contrast weighted by the lensing efficiency \cite{Kilbinger_2015}:
\begin{align} \kappa(\bm{\theta}) &= \frac{3\Omega_{\rm m}H_0^2}{2c^2} \int_0^{\infty} \dd\chi q_i(\chi) \frac{\delta_{\rm m}\left(\chi\bm{\theta}, \chi; z(\chi)\right)}{a(\chi)}, \label{eq:kappa-field} \end{align}
where $\Omega_{\rm m}$ and $H_0$ is the matter density and Hubble parameter at redshift zero, $c$ is the speed of light, $\chi$ is the comoving distance, $a(\chi)$ is the scale factor, and $\delta_{\rm m}$ is the matter density contrast. The lensing efficiency, $q_i(\chi)$, depends on the source galaxy distribution, $p(\chi)$:
\begin{align} q_i(\chi) &= \int_\chi^\infty \dd\chi' p_i(\chi') \frac{\chi'-\chi}{\chi'}, \quad \int \dd\chi p(\chi) = 1. \label{eq:lensing-efficiency} \end{align}

The weak lensing shear field in Cartesian coordinates is defined as $\gamma_c(\bm{\theta}) = \gamma_1(\bm{\theta}) + i\gamma_2(\bm{\theta})$, where $\gamma_1$ and $\gamma_2$ are shear components aligned with the x-axis and rotated by $45^\circ$, respectively. When projected onto a frame rotated by an angle $\zeta$, the spin-2 transformation gives:
\begin{align} \gamma(\bm{\theta}; \zeta) &= \gamma_{\rm t}(\bm{\theta}; \zeta) + i\gamma_\times(\bm{\theta}; \zeta) = -\gamma_{\rm c}(\bm{\theta}) e^{-2i\zeta}, \end{align}
where $\gamma_{\rm t}$ and $\gamma_{\times}$ are the tangential and cross components of the shear.

The Fourier transform of the shear field is defined as:
\begin{align} \gamma_{\rm c}(\bm{\theta}) &= \int \frac{\dd^2\bm{\ell}}{(2\pi)^2} \gamma_{\rm c}(\bm{\ell}) e^{-i\bm{\ell} \cdot \bm{\theta}}, \end{align}
with a similar relation for the convergence field. In Fourier space, the shear and convergence fields are related by:
\begin{align} \gamma_{\rm c}(\bm{\ell}) &= \kappa(\bm{\ell}) e^{2i\beta}, \label{eq:shear-convergence-relation} \end{align}
where $\beta$ is the polar angle of the Fourier mode $\bm{\ell}$.

Under the Limber approximation, we write the convergence power and bi-spectra as
\begin{equation}
P_{\kappa}(\ell) = \frac{9\Omega_{\rm m}^2H_0^4}{4c^4}\int_{0}^{\infty}\dd\chi\frac{q_i(\chi)q_j(\chi)}{a^2(\chi)}P_{\delta}\left(\frac{\ell}{\chi},z(\chi)\right),
\label{powercov}
\end{equation}
\begin{align}
\begin{split}
    B_\kappa(\ell_1, \ell_2, \ell_3) &= \frac{27\Omega_{\rm m}^3H_0^6}{8c^6}\int_{0}^{\infty}\dd\chi\frac{q_i(\chi)q_j(\chi)q_k(\chi)}{a(\chi)^3\chi}\\
    &\hspace{2em}\times B_{\delta}\left(\frac{\ell_1}{\chi},\frac{\ell_2}{\chi},\frac{\ell_3}{\chi},z(\chi)\right).
\end{split}
\label{biconv}
\end{align}

\subsection{Correlation functions}

The cosmic shear two-point and three-point correlation functions probe the matter distribution in real space. In order to theoretically model these observables, we start with the matter power spectrum and bispectrum. We model the first with the revised Halofit formula \cite{Takahashi_2012}, and the second with the BiHalofit formula \cite{Takahashi.Shirasaki.2019}. The convergence power and bi-spectrum are computed, next, from Eqs. \ref{powercov} and \ref{biconv}. 

By decomposing the convergence power spectrum into E and B modes, we can compute the shear two-point functions $\xi_{+}$ and $\xi_{-}$ by
\begin{align}
    \xi_{+}(\theta) = \int_0^{\infty} \frac{\ell d\ell}{2\pi} J_0(\ell\theta)[P_{\kappa}^E(\ell)+P_{\kappa}^B(\ell)],\\
    \xi_{-}(\theta) = \int_0^{\infty} \frac{\ell d\ell}{2\pi} J_4(\ell\theta)[P_{\kappa}^E(\ell)-P_{\kappa}^B(\ell)],
\end{align}
For the modeling of the shear three-point functions, we make use of the multipole formalism, as implemented in our \software{fastnc} code \cite{sugiyama2024fastmodelingshearthreepoint}. We use the natural components of the three-point function, as defined by \citet{Schneider.Lombardi.2002}, and write them in the $\times$-projection introduced by \citet{Porth.Schneider.2023}. We obtain

\begin{align}
\begin{split}
    \Gamma_0^{\times}(\theta_1, \theta_2, \phi) = 
    &-\int\frac{\dd^2\bm{\ell_1}}{(2\pi)^2}\frac{\dd^2\bm{\ell_2}}{(2\pi)^2}
    e^{-i\bm{\ell_1}\cdot\bm{\theta}_1-i\bm{\ell_2}\bm{\theta}_2} \\
    &\times
    B_\kappa(\ell_1, \ell_2, \alpha)
    e^{2i\sum_i\beta_i}
    e^{-3i(\varphi_1+\varphi_2)}, 
    \label{eq:Gamma0theo}
\end{split}
\end{align}
with similar expressions for $\Gamma_1$, $\Gamma_2$ and $\Gamma_3$. 

The bispectrum, now written in terms of two $\ell$ modes and the angle $\alpha$ between them, can be expressed as a sum over Legendre polynomials. Following \citet{sugiyama2024fastmodelingshearthreepoint}, we  compute a sequence of multipoles of the natural components of the 3PCF for each combination of two triangle sides $\theta_1$ and $\theta_2$. The full three-point function for $\theta_1$, $\theta_2$, and the opening angle $\phi$ is calculated next as a sum over all the computed multipoles.

\subsection{Mass aperture statistic}

In this work, we extract higher-order lensing information using the mass aperture statistic, a localized probe of the projected matter distribution. The skewness of the mass aperture, $\mapcube$, acts as a compressed summary of the non-Gaussian signal contained in the shear three-point function. 


The utility of $\mapcube$ as a compressed statistic has been demonstrated in recent work. As shown by \citet{Heydenreich.Schneider.2022}, this quantity captures nearly the full constraining power of the shear 3PCF, matching the performance of its leading principal components. This makes $\mapcube$ a practical and powerful summary for cosmological inference.

We obtain the mass aperture statistic from the natural components of the 3PCF following the procedure detailed by \citet{Jarvis.Jain.2003}. One advantage of using this procedure instead of computing it from a convolution of the bispectrum is that we maintain a consistent 3PCF binning effect between our theoretical model and our measurements. Finally, we run our full theoretical pipeline on a set of cosmologies taken from a Sobol sequence in parameter space \cite{sobol}. We sample over $\Omega_m$, $S_8$, $h_0$, $\Omega_b$, and $n_s$ in $\Lambda$CDM.

We use our results to train a neural network emulator for use in our cosmological inference pipeline. Our emulator computes the z-dependent mass aperture statistic $\mapcube (\theta, z)$. In order to extract our theory vector from it, we integrate the emulator predictions over the line-of-sight using the kernel $q(\chi)$, which includes both the lensing kernel and the intrinsic alignment kernel (See Section~\ref{sec:systematics}). This integration is given by
\begin{equation}
\mapcube(\theta)_{ijk} = \int \frac{d\chi}{\chi}\frac{q_i(\chi)q_j(\chi)q_k(\chi)}{a(\chi)^3} \mapcube(\theta, z(\chi)) 
\end{equation}
We use six hidden layers and a sequence of decreasing learning rates. We use a testing set of 171 cosmologies, each with 144 outputs, and find that 
the network error is below $0.29\%$ for $99\%$ of the samples,
and below $1.02\%$ for $100\%$ of the samples. We also build 
another emulator for $w$CDM,
for which we construct a new training set sampling over $w_0$ as an additional parameter. We find the error across the 262 testing set cosmologies to be less than $0.7\%$ for $99\%$ of the samples. The emulator building process and the final network hyperparameters, along with the ranges of the sampled parameters, are described in detail in our Paper~I.

\subsection{Modeling of systematics}

\label{sec:systematics}
Here, we briefly review the systematic parameters used in our analysis, referring the reader to our methodology work for the complete description. For redshift calibration and multiplicative shear biases, we follow the procedure of the fiducial two-point DES-Y3 analysis \cite{Secco.Weller.2022,Amon.Weller.2021,Myles_2021,MacCrann_2021}.

We account for the possibility of mean shifts in the redshift distributions by introducing four photo-z shift parameters $\Delta z_i$, one for each redshift bin $i$:
\begin{align}
    p_i(z) \rightarrow p_i(z-\Delta z_i).
\end{align}
Residual multiplicative biases on the shear are dealt with by introducing four parameters $m_i$, one for each redshift bin $i$. They are included in the correlation functions by
\begin{align}
    &\xi_\pm^{ij} \rightarrow (1+m_i)(1+m_j)\xi_\pm^{ij},\\
    &\Gamma_\mu^{ijk}\rightarrow (1+m_i)(1+m_j)(1+m_k)\Gamma_\mu^{ijk}.
\end{align}
We model intrinsic alignment through the non-linear alignment (NLA) paradigm, which yields us the free parameters $A_1$ (amplitude parameter) and $\alpha_1$ (for redshift evolution). While more complex models have been studied in the context of two-point functions, we aim for a consistent modeling between second- and third-order correlations, as suggested by \citet{Pyne.Joachimi.2022}, and so we adopt NLA both for 2PCF and 3PCF. Our NLA modeling is implemented by modulating the lensing kernel through \cite{Gatti.collaboration.2020,Krause.Weller.2017}
\begin{align}
    q_i(\chi) \rightarrow q_i(\chi) + f_{\rm IA}\left(z(\chi)\right) p_i(\chi)\frac{\dd z}{\dd \chi},
    \label{align1}
\end{align}
where $f_{\rm IA}(z)$ is given in terms of the intrinsic alignment (IA) free parameters by
\begin{align}
    f_{\rm IA}(z) = - A_1 \left(\frac{1+z}{1+z_0}\right)^{\alpha_1} \frac{c_1 \rho_{\rm crit}\Omega_{\rm m,0}}{D(z)}.
    \label{eq:f-IA}
\end{align}
Here, we use $z_0=0.62$ as the pivot redshift (this is a matter of convention, we follow the choice of \citet{Secco.To.2021}). Our growth function $D(z)$ is normalized to unity at redshift zero, and $c_1\rho_{\rm crit}=0.0134$ is a constant determined through SuperCOSMOS observations \cite{Bridle_2007}.

The question of whether such NLA modeling is sufficient for a Y3 analysis has been addressed by \citet{Secco.To.2021}. While the fiducial Y3 cosmic shear analysis makes use of the tidal alignment and tidal torquing (TATT) model \cite{Blazek.2019}, the performance of both models is compared, and the shift in constraints is found not to be significant. The IA signal is found to be low, and an analysis of statistical model selection finds a preference for models with fewer parameters. TATT is a more realistic model, but it also adds complexity in parameter space, which can lead to the degradation of constraints. While improving our IA model will be necessary for larger datasets, we leave the complete development of a TATT model of the 3PCF and the mass aperture statistic for future work. Advancements in this direction are being made, with a theoretical modeling of the IA bispectra established by \citet{bakx1} and validated against simulations of dark matter halo shapes \cite{bakx2}.

\section{Measurements}\label{sec:measurements}

\begin{figure*}
    \centering
     \includegraphics[width=\linewidth]{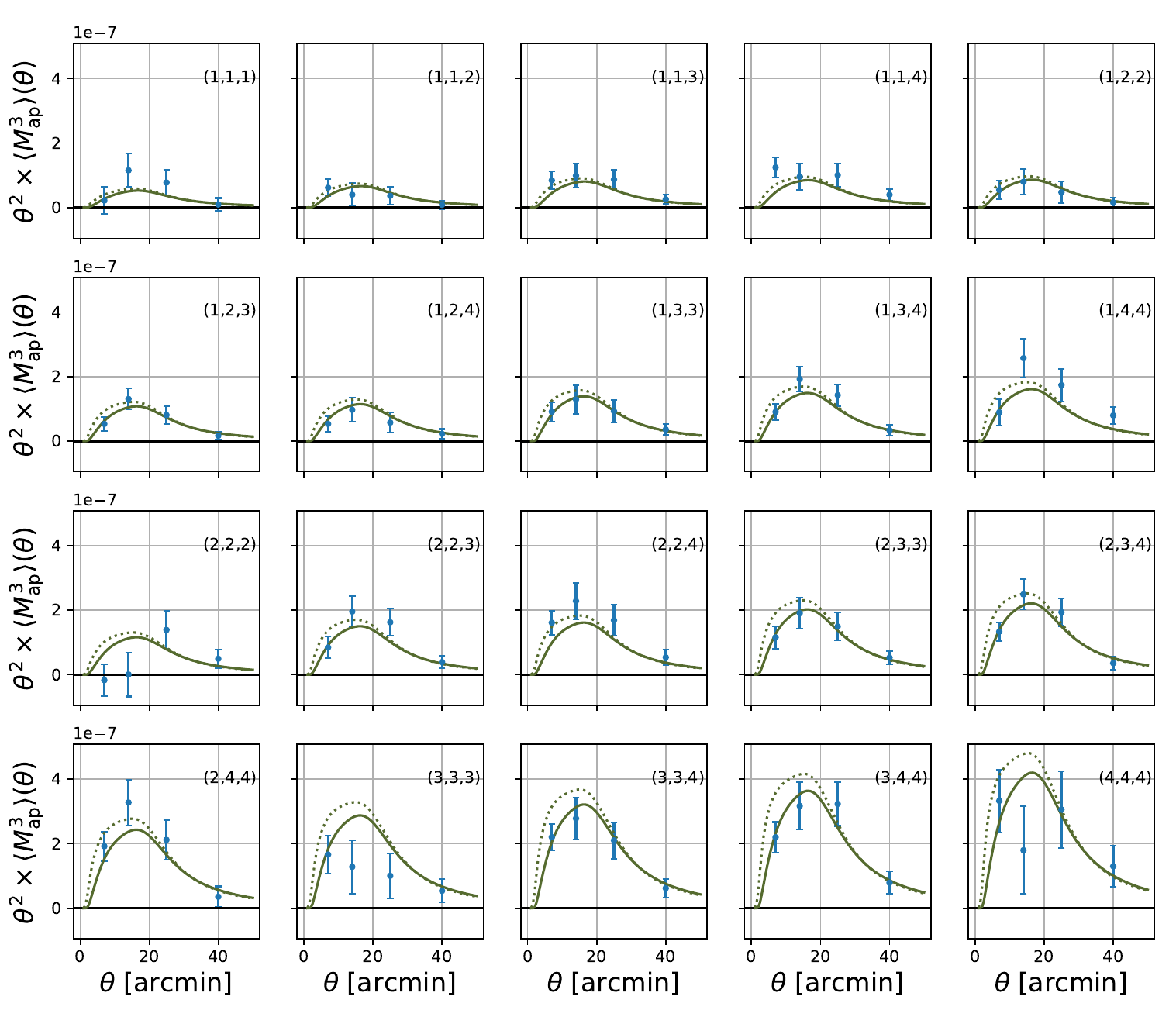}
    \caption{The third moment of the mass aperture statistic Y3 data vector.  The panels show the aperture mass statistic as a function of filter radii $\theta$ for different redshift-bin combinations $(i,j,k)$ indicated on the upper right corner of each panel. The error bars are estimated from our simulation-based covariance. The theoretical prediction at our joint best fit cosmology, as described in Section \ref{sec:result}, is shown by the continuous green line, which indicates the prediction with our 3PCF scale cut at 8 arcmin. The green dotted line shows the theoretical prediction with no scale cuts on the 3PCF. This is not expected to match the data (which does have scale cuts applied) but it shows that only a modest amount of the signal is lost by the scale cuts (difference between continuous and dotted lines).}
    \label{fig:meas}
\end{figure*}

\begin{figure*}
    \centering
    \includegraphics[width=\linewidth]{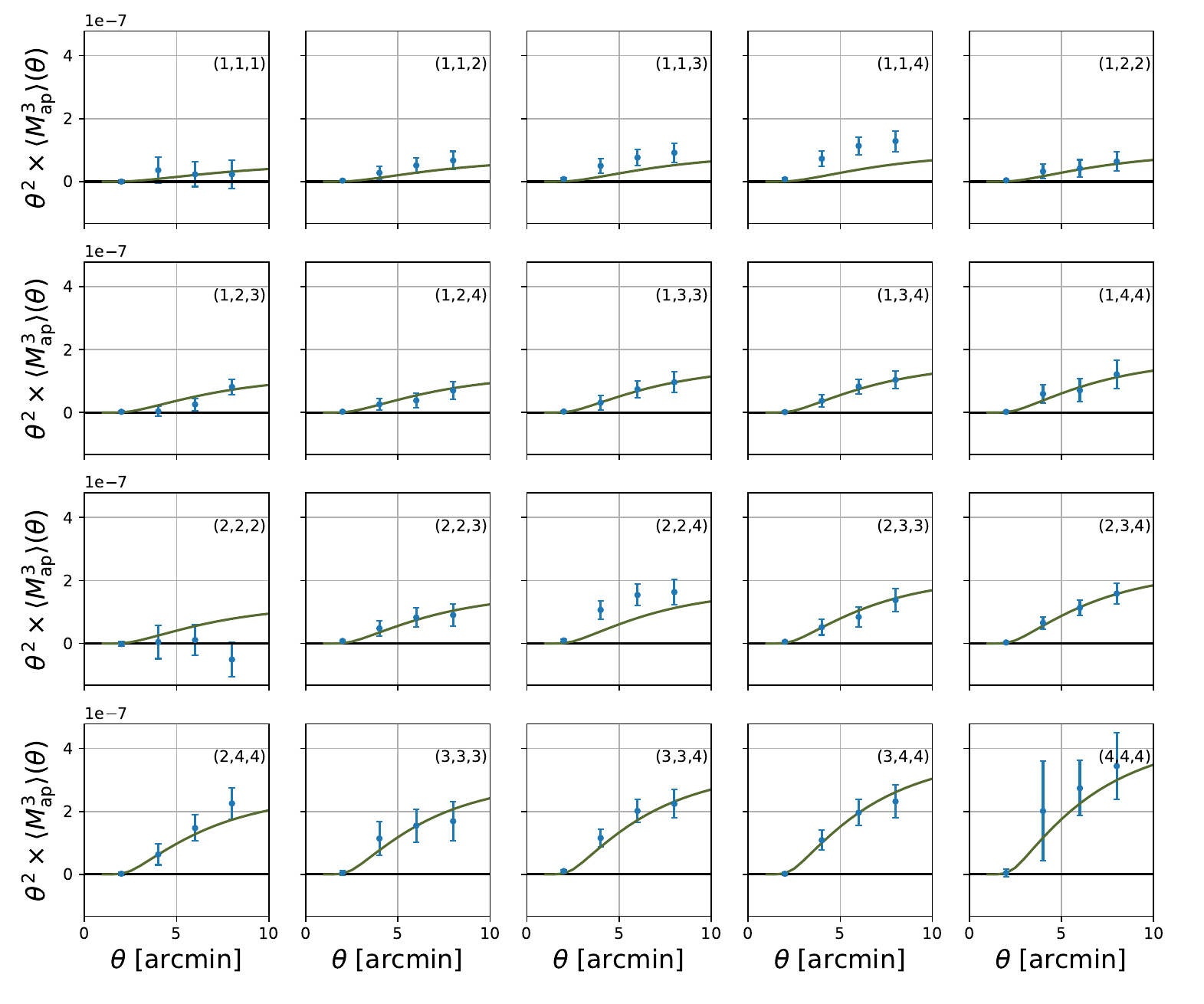}
    \caption{Mass aperture statistic Y3 data, as in Figure~\ref{fig:meas}, but with aperture filters at small scales. The green continuous line shows the theoretical prediction at best fit cosmology with our 3PCF scale cut at 8 arcmin. The dark-matter-only theoretical prediction is an adequate fit to the small scale data, with no systematic departure present across most of the redshift bin combinations.}
    \label{fig:meas2}
\end{figure*}

\subsection{Three-point correlation function and mass aperture statistic}
The three-point correlation function has been previously measured on the DES-Y3 cosmic shear catalog by \cite{Secco.Weller.2022}. Their measurements were performed in two different sets: (i) one non-tomographic measurement over the whole catalog, and (ii) tomographic measurements of auto and cross-correlations between two newly defined redshift bins, created by merging Y3 source bins 1 and 2, and also merging bins 3 and 4.

We perform a new set of measurements of the cosmic shear 3PCF on DES Y3 data in order to complete a full tomographic measurement and prepare our $\mapcube$ data vector for parameter inference. We use the METACALIBRATION weak lensing shape catalog described by \cite{Gatti.Wilkinson.2020}. The relation between the measured ellipticity $(e_1,e_2)$ and the cosmic shear $(\gamma_1,\gamma_2)$ is described by the response matrix $\textbf{R}_{\gamma}$. The METACALIBRATION algorithm artificially applies an additional shear to the original images, creating four additional versions of the ellipticity catalog: one positively sheared and one negatively sheared version for each $e_i$ component. Using finite differences between these catalogs, the response matrix is estimated by
\begin{equation}
\textbf{R}_{\gamma, ij} = \frac{e_i^+-e_i^-}{\Delta \gamma_j}.
\end{equation}
To the shear response matrix, we add the selection response $\textbf{R}_S$ matrix, as defined by \cite{Sheldon.Huff.2017}, to account for selection effects. We use as our final response matrix $\textbf{R} = \textbf{R}_{\gamma}+\textbf{R}_S$.

Following \cite{Secco.Weller.2022}, we compute our three-point statistic using the unsheared ellipticity map from METACALIBRATION, masking it separately for each of the four DES Y3 source redshift bins. We subtract from the ellipticity component $e_i$ of each galaxy its mean over the whole masked catalog, and divide it by the scalar $R = (\textbf{R}_{11}+\textbf{R}_{22})/2$, taking it as representative of the response matrix $\textbf{R}$.

We perform the measurements of the three-point correlation function with \software{TreeCorr}\cite{Jarvis.Jain.2003}, using the Multipole binning scheme. In this scheme, each triangle configuration is described by two side lengths and a set of multipoles of the opening angle. We take as the maximum multipole $\mathrm{max}_{n}=100$. For the side binning, we take 20 bins logarithmically spaced between $\theta_{\rm min}=0.5'$ and $\theta_{\rm max}=80'$. The computational runtime of the measurement for each redshift bin combination is of the order of 3 hours with 128 CPUs at the Perlmutter system at the National Energy Research Scientific Computing Center (NERSC). The Multipole binning scheme, following the proposal by \citet{Porth.Schneider.2023}, allows for a measurement that scales linearly with the number of galaxies.

Next, we convert these measurements into the side-angle-side binning scheme, using 63 bins for the opening angle, which go from 0 to $\pi$ with a linear spacing of $\phi=0.05$. 

Then, we perform a scale cut at the level of the 3PCF, removing all triangle contributions with one or two sides smaller than $\theta = 8'$. This cut is motivated by considering the resolution of the simulations from which we compute our covariance matrix, and is designed so that we have a robust covariance across all scales.

Finally, these measurements are transformed into $\mapcube$ data. The expression of $\mapcube$ as an integral of the natural components $\Gamma_i$ was developed by \citet{Jarvis.Jain.2003}, and is implemented in \software{TreeCorr} through matrix multiplication of the binned 3PCF measurements.

Through this procedure, the mass aperture statistic is computed from the cosmic shear catalog, without the need of any convergence map reconstruction. There is another way to measure $\mapcube$ from the shape catalog through reconstructing the convergence map, which, however, suffers from masking effects on the convergence map and mass aperture field. Therefore, our approach of measuring $\mapcube$ directly from the shape catalog, computing it as an integral of the 3PCF, naturally
circumvents biases introduced by masked regions or irregular edges of survey footprint \citep{Secco.Weller.2022}, and enables robust comparison of the theoretical prediction of $\mapcube$.

We use four equal-aperture filters, at $7'$, $14'$, $25'$, and $40'$. We introduce our smallest filter at $7'$ in order to avoid potential baryonic contamination. Our largest filter is chosen to be at $40'$, 
because we do not expect much gain in signal-to-noise ratio, and to avoid observational systematics at the large scales.

Our set of measurements is presented in Figure \ref{fig:meas}. The total $\mapcube$ signal-to-noise, computed using the simulation-based covariance, is 12.7. The largest contributions towards the total S/N ratio are the bin combinations 234, 334, and 344, each with individual S/N ratios higher than 6. The lower redshift bins, however, still carry contributions, with the total $\mapcube$ S/N being reduced to 10.5 if we remove all data that includes the first redshift bin.

We also measure the $\mapcube$ with four additional filters at $2'$, $4'$, $6'$ and $8'$. We do not include these data points in our fits, but instead use them to further validate our methodology, by comparing them to the dark-matter-only theoretical prediction at our best-fit cosmology presented in Section \ref{sec:result}. We show these measurements in Figure \ref{fig:meas2}.

\begin{figure*}
    \centering  
    \includegraphics[width=0.48\textwidth]{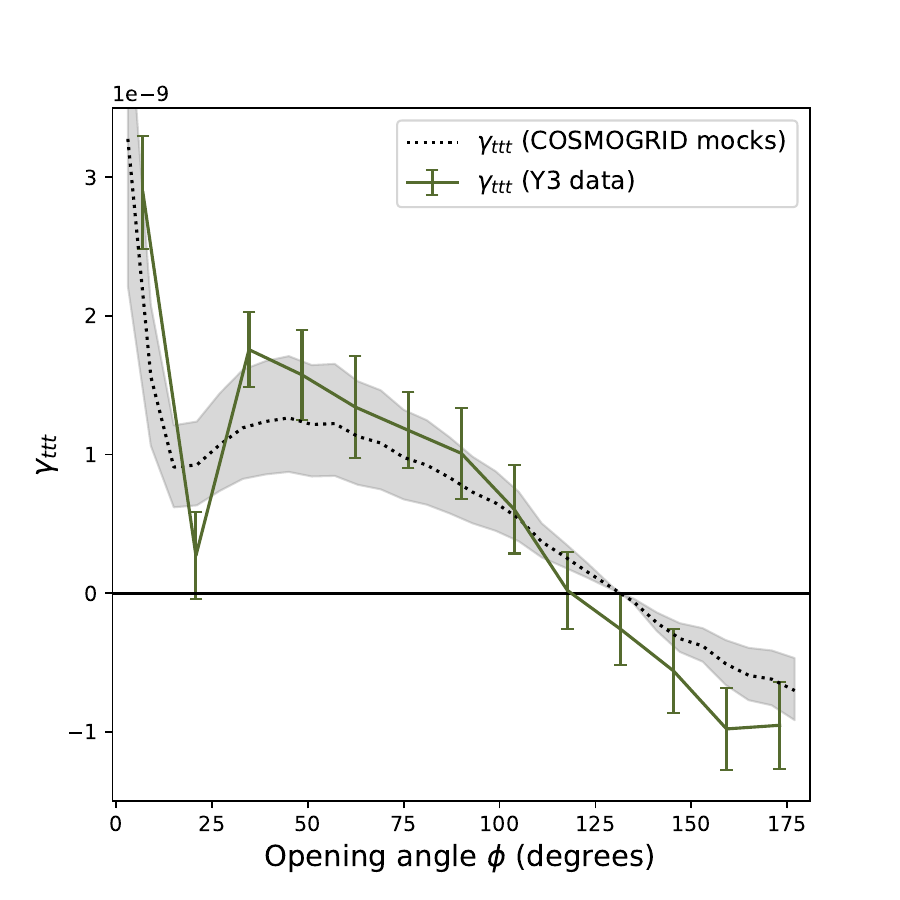}
    \includegraphics[width=0.48\textwidth]{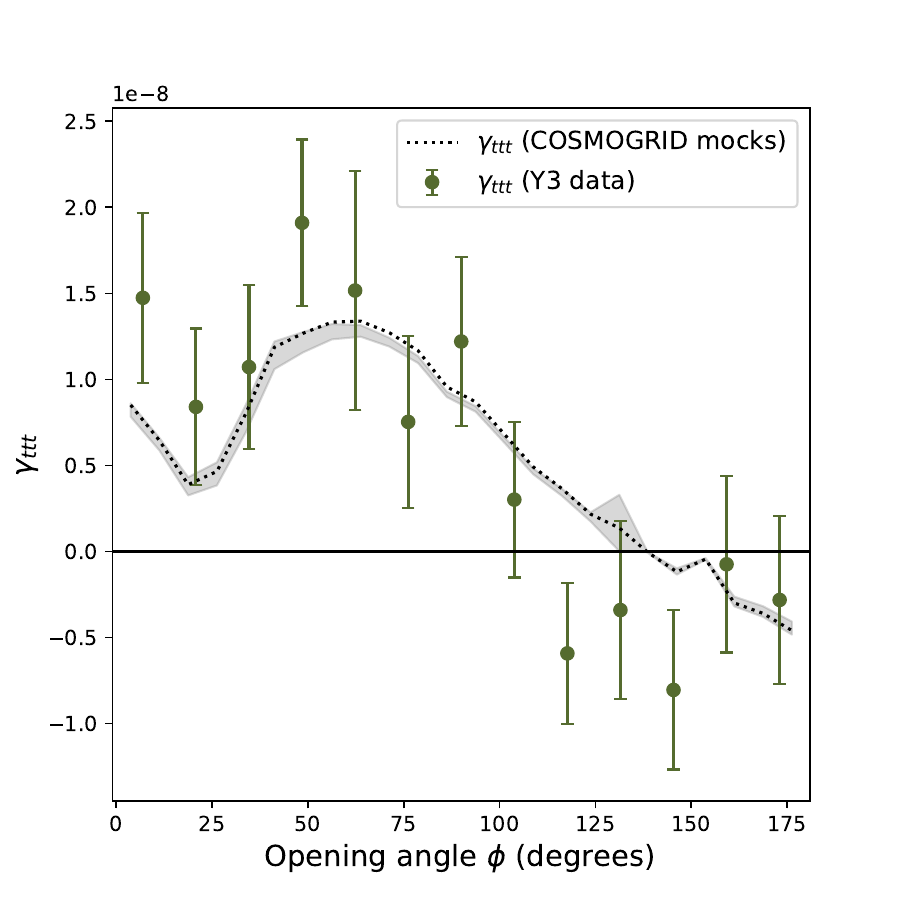}
    \caption{Measurement of the isosceles three-point correlation function on DES Y3 data. We performed the measurement without redshift tomography, and present the results for scales below 18 arcmin (right panel) and between 18 and 80 arcmin (left panel). Both panels show the characteristic shape and zero crossing of the shear 3PCF. The error bars are estimated from the jackknife covariance. To provide a comparison with theoretical expectations, the dotted curve shows  our simulation measurements, rescaled to account for the difference between cosmological parameters.   The $1\sigma$ intervals around our $\mapcube$ best fit cosmology is shown by the gray shaded region. 
    }
    \label{isosc_meas_1}
\end{figure*}

Finally, we also perform a non-tomographic measurement of the three-point correlation function over the whole catalog at the same scales of $\theta_{
\rm min}=0.5'$ and $\theta_{\rm max}=80'$. We use this measurement to study the configuration dependence of the 3PCF. We select near-isosceles triangles, defined as those for which the $d_2$ angular scale bin equals that of $d_3$. We then perform a weighted average over $d_2$ and $ d_3$, splitting our measurement between small scales ($d_2 \approx d_3< 18'$) and larger scales ($18' < d_2 \approx d_3< 80'$). We show our results in Figure \ref{isosc_meas_1}. We take our error bars from jackknife sampling of the Y3 data, which we divide into 100 patches. We select this method instead of taking our covariance from simulations in order to avoid small scale biasing from simulation resolution. For the small scale selection, our S/N is 6.9, while for the larger scales, we find S/N = 11.1. We show the signal together with a reference theoretical expectation range, which we compute by averaging the non-tomographic signal of a set of CosmoGridV1 realizations. The amplitude of the signal is rescaled by the average ratio between the expected signals at CosmoGridV1 cosmology and at our $\mapcube$ best fit cosmology. The upper and lower limits of our range are computed from cosmologies which are $1\sigma$ away from our best fit.

When we compare our measurements to those from \citet{Secco.Weller.2022}, we notice that both our Y3 and CosmoGridV1 measurements remain shifted towards negative values as the opening angle approaches 180$\deg$. In the case of the larger scales, we can see clearly that there is no minimum point near such angles. The feature identified by \citet{Secco.Weller.2022} of a rise in the signal near $\phi=180\deg$, which doesn't match theoretical expectations, can be explained as the result of uneven binning when using the \software{ruv} binning option in \software{TreeCorr}. We remove this effect by performing our analysis with the \software{multipole} binning scheme, which adapts the multipole-based estimator developed by \cite{Porth.Schneider.2023} into the tree structure.

The shape dependence of the 3PCF carries interesting features that can potentially serve as null tests for systematics. One such example is the zero crossing at an opening angle around $\phi\sim120\deg$. The expected sign of different components of the shear three-point function was studied by \citet{Zaldarriaga.Scoccimaro}. For isosceles triangles close to an equilateral configuration, we expect a tangential shear on the three vertices of the triangle. When the isosceles triangle approaches a flattened configuration, the presence of mass in the region inside the triangle induces almost always a radial shear on the obtuse angle vertex, and a tangential shear on the small angle vertices. This leads to a negative tangential $\gamma_{ttt}$. 
 
Between these two limiting cases, equilateral and flattened isosceles triangles, the isosceles cosmic shear signal should always have a zero crossing.
The angle of the zero-crossing is determined by the geometrical configuration of the mass and the triangle on the sky as described above, while the observational systematics, PSF systematics, does not follow this picture and can produce a shift in the position of zero-crossing. Therefore, the location of the zero-crossing could be used as a null test of the systematics. We leave this potential use of the zero-crossing as a null test for future research.

\section{Methodology}\label{sec:methods}

\subsection{Parameter inference pipeline}
To infer cosmological and nuisance parameters from the DES Y3 data, we adopt a Bayesian framework in which the posterior distribution of the parameters $\bm{p}$ given a data vector $\bm{d}$ is proportional to the product of the likelihood and the prior distribution:

\begin{align} \mathcal{P}(\bm{p}|\bm{d}) \propto \mathcal{L}(\bm{d}|\bm{p}) \Pi(\bm{p}). \end{align}
We adopt the likelihood derived in \citet{Percival.Friedrich.2021}, which naturally yields posterior credible intervals compatible with those from the frequentist approach. The log-likelihood is expressed in terms of the $\chi^2$ statistic that measures the discrepancy between the observed data and the theoretical model prediction $\bm{t}(\bm{p})$. Thus, we have
\begin{equation} \ln \mathcal{L}(\bm{d}|\bm{p}) = -\frac{m}{2} \ln \left(1 + \frac{\chi^2}{N_{\text{real}} - 1}\right) + \text{const}, \end{equation} \begin{equation} \chi^2 = [\bm{d} - \bm{t}(\bm{p})]^\mathrm{T} \bm{C}^{-1} [\bm{d} - \bm{t}(\bm{p})],
\label{eq:percival}
\end{equation}
where $N_{\text{real}} = 796$ is the number of \textsc{CosmoGridV1} realizations used to estimate the covariance matrix. We also define $N_{\bm{p}}$ as the number of free parameters and $N_{\bm{d}}$ as the length of the data vector. Then, we  can write the $m$ factor as
\begin{equation} m = N_{\bm{p}} + 2 + \frac{N_{\text{real}} - 1 + f_D}{1 + f_D},
\end{equation}
where $f_D$ is the Dodelson-Schneider factor \cite{Dodelson.Schneider.2013}, given by
\begin{equation} f_D = \frac{(N_{\bm{d}} - N_{\bm{p}})(N_{\text{real}} - N_{\bm{d}} - 2)}{(N_{\text{real}} - N_{\bm{d}} - 4)(N_{\text{real}} - N_{\bm{d}} - 1)}.\end{equation}
Our fiducial data vector combines both two-point and three-point statistics: it includes the $\xi_+$ and $\xi_-$ shear correlation functions as well as the third-order mass aperture statistic, $\langle \mathcal{M}_{\rm ap}^3 \rangle$. For $\xi_{\pm}$, we use 20 angular bins logarithmically spaced between $2.5'$ and $250'$ for each redshift bin pair. We apply the fiducial Y3 scale cuts, described by \citet{Krause.Fang.2021}, to mitigate the effects of baryonic physics, retaining 227 data points in total. We apply MOPED compression \cite{Heavens.Vianello.2017} on the $\xi_{\pm}$ section of our data vector, following the procedure described in Paper~I. We use the analytical $\xi_{\pm}$ covariance matrix to determine the MOPED compression matrix, and then use the simulated covariance for cosmological inference. We found that this mitigates the risk for overconfidence that would be present if the compression matrix had been derived from a simulated covariance.

For the three-point part, we use all 80 mass aperture data points, spanning 20 redshift combinations and four aperture radii from $7'$ to $40'$. This choice of scales is described and shown to be robust against baryonic feedback in Paper I. In our simulated analysis, we find that through this set of filters the small scale baryonic suppression of the bispectrum is canceled out by the mid scale enhancement, generating a small net enhancement effect. A joint analysis of $\xi_{\pm}$ and $\mapcube$, therefore, becomes even more robust to the effect of baryons.

We consider both $\Lambda$CDM and $w$CDM models in our inference process. For $\Lambda$CDM, our parameter set includes the six base parameters ($\Omega_m$, $S_8$, $h_0$, $\Omega_b$, $m_{\nu}$, and $n_s$) and 10 nuisance parameters to account for the effects described in Section~\ref{sec:systematics}. These nuisance parameters are four photo-z shift parameters $\Delta z_i$ and four multiplicative shear biases $m_i$, corresponding to the four tomographic redshift bins, plus the two intrinsic alignment parameters $A_1$ and $\alpha_1$. For $w$CDM, we allow the dark energy equation of state parameter $w_0$ to vary, resulting in one additional free parameter. Our priors on the cosmological and nuisance parameters are the same as those of our simulated analysis and are shown in Paper I.

Our pipeline is implemented in \textsc{CosmoSIS} \cite{Zuntz_2015}, which provides modular support for likelihood evaluation and sampling. We use its standard modules for modeling the two-point function and developed custom modules for computing the mass aperture three-point function, based on the \textsc{fastnc} framework introduced by \citet{sugiyama2024fastmodelingshearthreepoint}.

Posterior sampling is performed using the \textsc{MultiNest} algorithm \cite{Feroz.Bridges.2009}, integrated via the \textsc{CosmoSIS} interface. We perform our runs with \texttt{nlive} = 500, \texttt{efficiency} = 0.3, and \texttt{tolerance} = 0.1. 

We measure our level of improvement on the $\Omega_{\rm m}$ and $S_8$ constraints when adding $\mapcube$ to $\xi_{\pm}$ by comparing the figure-of-merit (FoM), which is defined by
\begin{equation}
    \text{FoM} = \frac{1}{\sqrt{\text{Cov}(\Omega_{\rm m}, S_8)}},
\end{equation}
where $\text{Cov}(\Omega_{\rm m}, S_8)$ is the covariance of the posterior samples in the $\Omega_{\rm m}$ and $S_8$ 2D parameter space after marginalizing over all the other sampled parameters.

\begin{figure*}
    \centering
    \includegraphics[width=\linewidth]{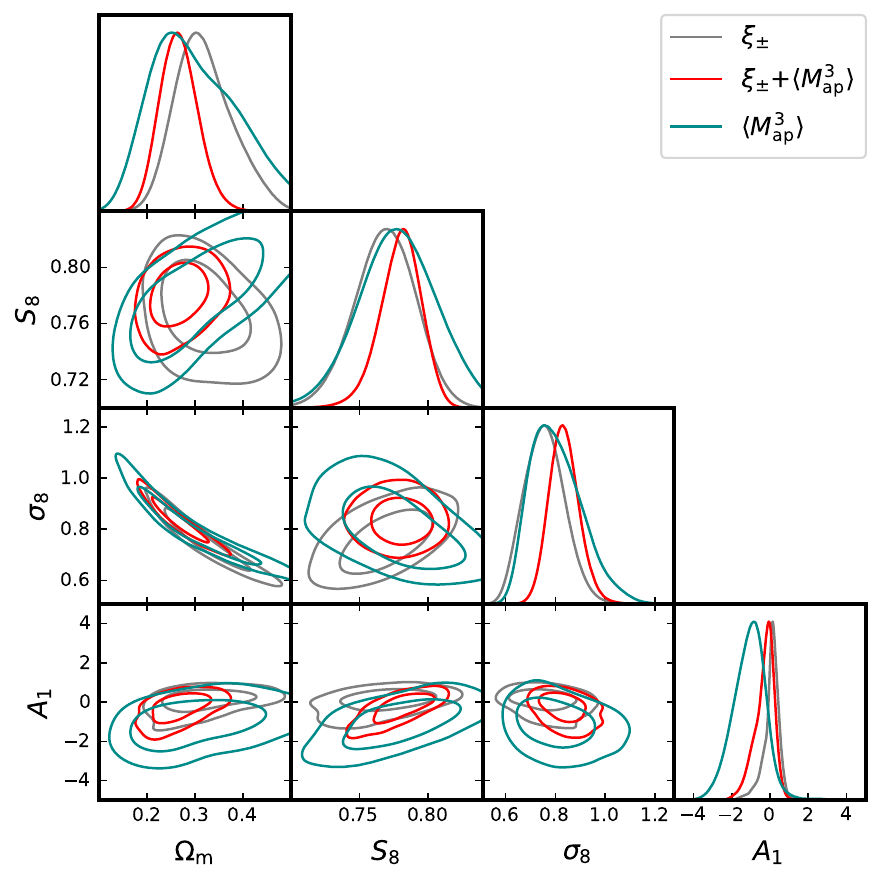}
    \caption{Parameter constraints from DES-Y3 data from separate and joint analyses of $\xi_{\pm}$ and $\mapcube$ within $\Lambda$CDM. The cosmological parameters shift less then 1$\sigma$ when moving from $\xi_{\pm}$ to a joint 2PCF and 3PCF analysis. We find an improvement of 111\% in the $\Omega_m$-$S_8$ figure-of-merit. This improvement is driven in part by the difference between the degeneracy directions for $\xi_{\pm}$ alone and $\mapcube$ alone. Here $A_1$ is the intrinsic alignment amplitude (see Eq. \ref{eq:f-IA})}
    \label{fig:y3result_full}
\end{figure*}

 \begin{figure}
    \centering
    \includegraphics[width=\linewidth]{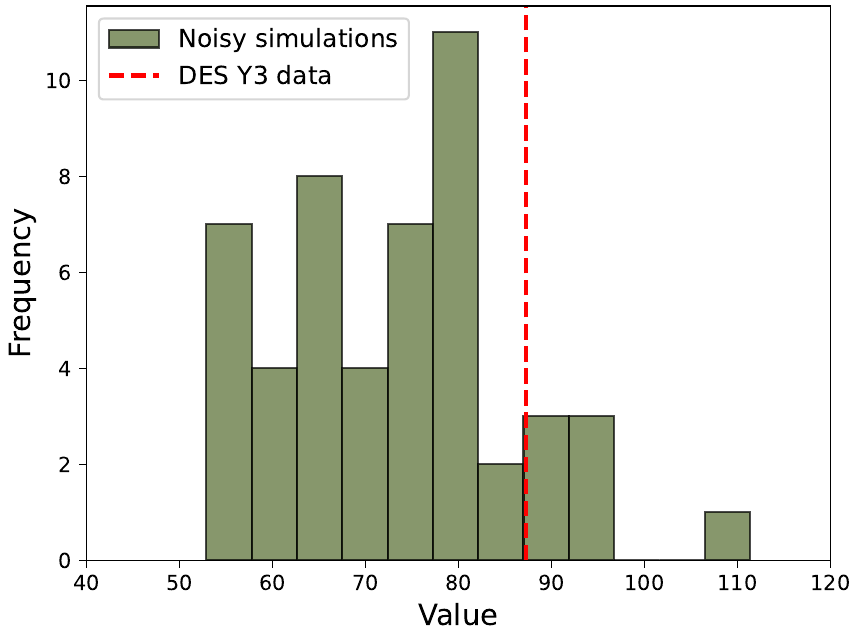}
    \caption{The $\chi^2$ distribution estimated from noisy mock simulations for the DES joint $\xi_{\pm}$+$\mapcube$ analysis. The value from the analysis with Y3 data is represented by the dashed red line. Our data $\chi^2$ is consistent with its expected distribution, with a probability of 14\% of a random realization having higher $\chi^2$. Our distribution average is 73.3, which is sufficiently close to our effective number of degrees of freedom ($N=89$).}
    \label{fig:chi2dist_unblind}
\end{figure}

\begin{figure*}
    \centering
    \includegraphics[width=0.8\linewidth]{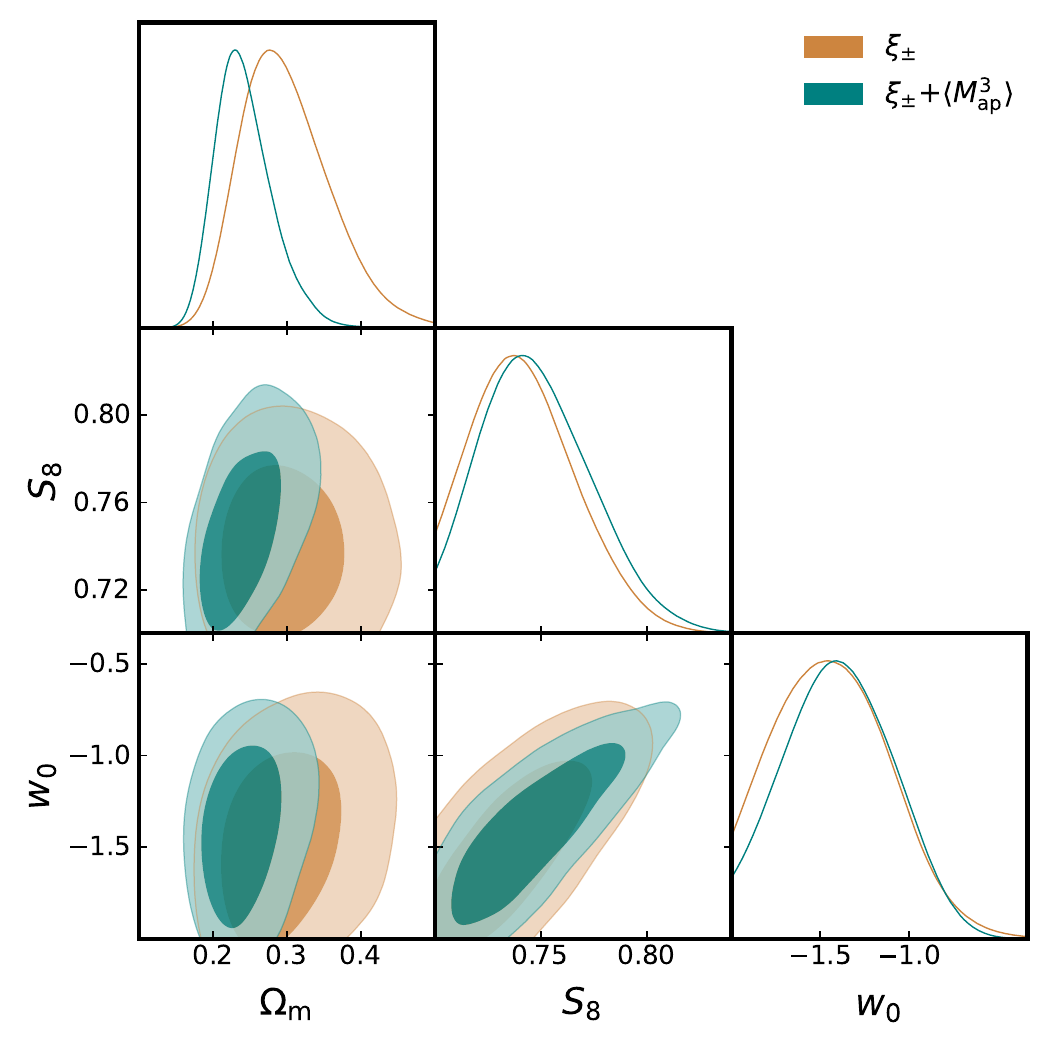}
    \caption{Parameter constraints from DES Y3 data using $\xi_{\pm}$ and $\mapcube$, as in Figure~\ref{fig:y3result_full} but for the $w$CDM model. We find an improvement of 66\% in the $\Omega_m$-$S_8$ figure-of-merit. Our improvement on the $w_0$ parameter is only 5\%, while the improvement on the joint $w_0$-$S_8$ constraint is 22\%. 
    }
    \label{fig:y3wcdm}
\end{figure*}

\begin{figure*}
    \centering
    \includegraphics[width=0.8\linewidth]{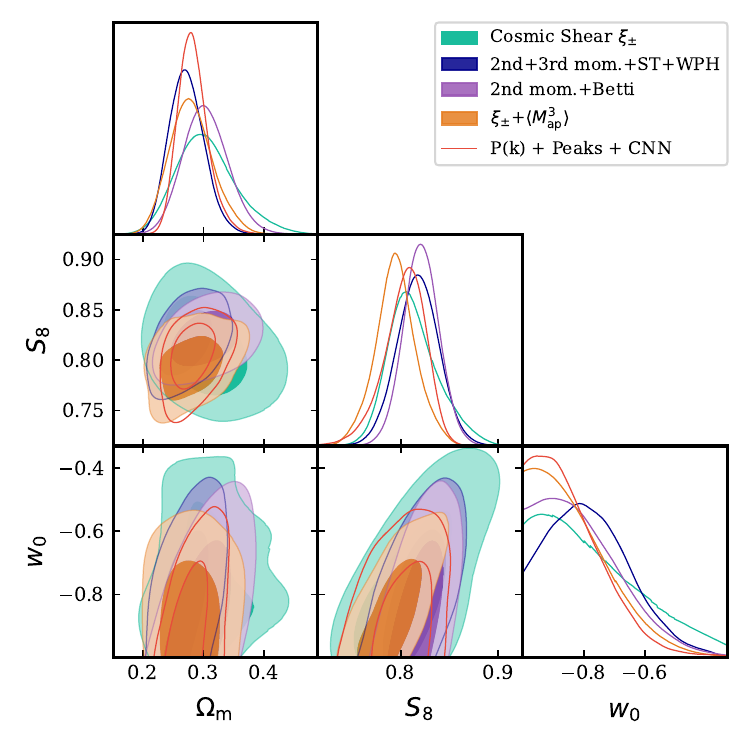}
    \caption{Parameter constraints from DES Y3 data using different higher-order statistics. We compare constraints by \citet{Gatti.Campailla.2024} from the second and third moment of the weak lensing mass map, scattering transforms, and wavelet phase harmonics, constraints by \citet{Prat.Gatti.Doux.2025} from the second moment and Betti numbers, constraints by \citet{Jeffrey.Whiteway.Gatti.2024} from power spectra, peak statistics, and CNN map-level inference, and the constraints from $\xi_{\pm}$ and $\mapcube$ presented in this work. We also include, as a baseline, the $\xi_{\pm}$ result by \citet{Prat.Gatti.Doux.2025}. We adapt the priors on our analysis to match those of the SBI-based statistics. All higher-order-statistics yield significant improvements relative to $\xi_{\pm}$ alone and are consistent with each other. 
    }
    \label{fig:HOS}
\end{figure*}

 \begin{figure*}
    \centering
    \includegraphics[width=\linewidth]{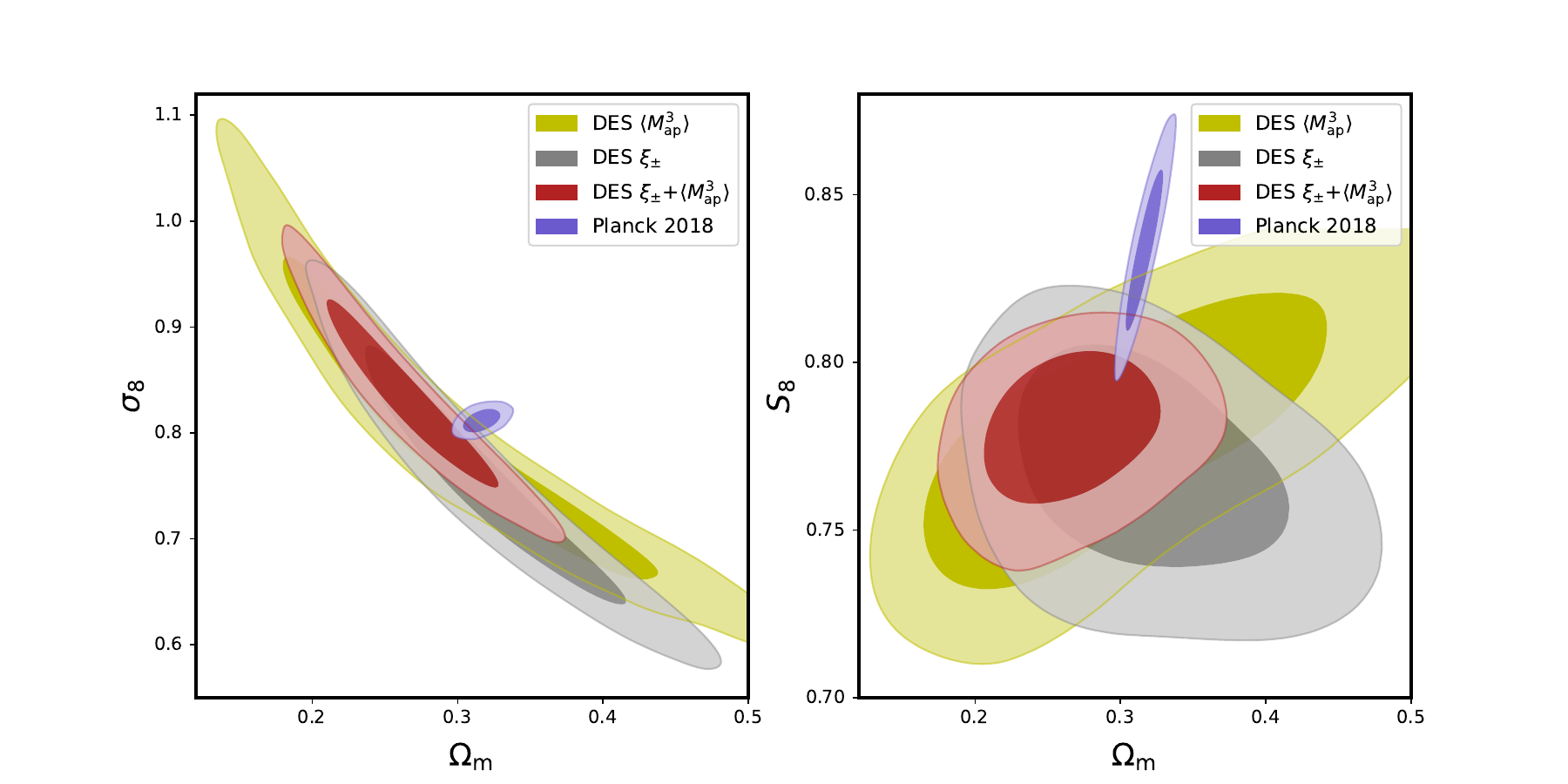}
    \caption{Tension between DES Y3 results from $\xi_{\pm}$ and $\mapcube$ and CMB results from Planck 2018. We use their TTTEEE+lowl/lowE results, which include temperature (TT), polarization E-modes (EE), the cross spectra (TE), and the low multipole values for both temperature and polarization. The $S_8-\Omega_m$ tension between the DES Y3 shear two-point analysis and Planck data is of $2.3\sigma$. While the tension is smaller for $\mapcube$ ($1.9\sigma$) due to it being less informative, the tension between the joint $\xi_{\pm}$ -- $\mapcube$ DES Y3 constraint and Planck remains at $2.3\sigma$.}
    \label{fig:s8vsplanck}
\end{figure*}

\subsection{Blinding}

It has been recognized that cosmological analyses are susceptible to the influence of biases from the one undertaking the analysis \cite{Brieden_2020} \cite{Novell-Masot}. The risk is that knowledge of results from other data sets can exert an influence over analysis choices. In order to avoid this risk, it is a common practice to perform data blinding while the pipeline is being tested. After consistency tests are performed on blinded data, the pipeline is then applied to the actual measured data. 

We perform blinding at two different levels \cite{Muir_2020}. Firstly, we perform a catalog-level blinding. We follow \citet{Gatti.Wilkinson.2020} and measure the two-point and three-point functions from a catalog with shifted ellipticities. Secondly, we also perform blinding at the level of the cosmological parameters. We hide the parameter values when displaying our posterior distributions and only view the relative values between our different data sets. We remove the blinding after ensuring that the $\chi^2$, the shift in cosmological parameter space, and the $\Omega_m$-$S_8$ figure-of-merit improvement, all fall within their expected distribution ranges extracted from a set of 50 noisy realizations. These tests are detailed in Appendix \ref{sec:appendix-blind}.

\section{Results}\label{sec:result}

\subsection{$\Lambda$CDM Cosmological Constraints}\label{sec:result-des-y3}

We use our pipeline on the Y3 shear catalog and compare the constraints from the two-point function alone, with those including three-point data. For the nominal $\Lambda$CDM model, these results are presented in Figure \ref{fig:y3result_full}. The improvement on the FoM on the $\Omega_m$-$S_8$ plane is $111\%$. This value is slightly higher than the one forecast by our methodology paper. We verify the reasonable range of expected FoM improvements as part of our pre-unblinding tests, described in Appendix \ref{sec:appendix-blind}.

Our cosmological constraints from the joint chain are
\begin{align}
\text{$\xi_\pm$+$\mapcube$:}
\begin{cases}
    \Omega_m &= 0.266^{+0.039}_{-0.040}\\
    S_8 &= 0.780\pm 0.015
\end{cases}.
\end{align}
We compare these results with those from our two-point analysis on Table \ref{tab:param_values}.

\begin{table*}[t]
    \caption{Mean of parameter posteriors from 2PCF and 2PCF+$\mapcube$ analysis.}
    \centering
    \setlength{\tabcolsep}{15pt}
    \renewcommand{\arraystretch}{1.5}
    \begin{ruledtabular}
    \begin{tabular}{lllll}
    Parameter & $\xi_{\pm}$ & $\xi_{\pm}$ + $\mapcube$ & $\xi_{\pm}$ ($w$CDM) & $\xi_{\pm}$ + $\mapcube$ ($w$CDM)\\ \hline
    $\Omega_{\rm m}$ &  $0.318^{+0.063}_{-0.059}$ & $0.266^{+0.039}_{-0.040}$  & $0.298^{+0.061}_{-0.059}$  & $0.242^{+0.038}_{-0.037}$ \\
    $S_8$ & $0.769\pm 0.022$ & $0.780\pm 0.015$ & $0.743\pm 0.025$  &  $0.749^{+0.027}_{-0.026}$\\ $\sigma_8$ & $0.757^{+0.080}_{-0.084}$ & $0.835^{+0.058}_{-0.060}$ & $0.756\pm 0.079$ & $0.841\pm 0.057$ \\ 
     $w_0$ & - & -  & $-1.42^{+0.32}_{-0.33}$ &$-1.39\pm 0.31$  \\
    \end{tabular}
    \end{ruledtabular}
    \label{tab:param_values}
\end{table*}

The consistency of the two-point and mass aperture data vectors can be verified by computing the shifts on parameter posteriors when using different parts of the data. Because $\mapcube$ alone has little constraining power, we compare constraints from $\xi_{\pm}$ directly with those from the joint analysis. Therefore, we compute the shift of the two-dimensional posterior peaks in the $\Omega_m$-$S_8$ plane when moving from $\xi_{\pm}$ to a joint analysis with $\mapcube$. We obtain a shift of around $0.5\sigma$, which falls within the expected statistical scatter from noisy simulations, with a p-value of 0.78. (see Appendix \ref{sec:appendix-blind} for blind tests). 

We evaluate the goodness-of-fit of our joint analysis by computing the $\chi^2$ at our maximum a-posteriori cosmology. We verify the $\chi^2$ of the joint best fit to be of $\chi^2=87.3$. Our number of degrees of freedom equals $N_d-N_{\text{eff}}$, where $N_d$ is our data vector length and $N_{\text{eff}}$ is an effective value which is related to the number of free parameters and to constraints from our priors. \citet{Raveri.Hu.2019} derive the expression for this value to be $N_{\text{eff}} = N_p - \text{tr}(C_{\Pi}^{-1}C_p)$, where $C_{\Pi}$ and $C_p$ are the parameter covariances of the prior and posterior distributions. We estimate
$N_{\text{eff}} \approx 7$, which gives  the number of degrees of freedom: $96-7=89$. Thus, our reduced $\chi^2$ is 0.98, which indicates a good fit to the model.

A second goodness-of-fit test was also performed by running our pipeline on a set of 50 noisy realizations, generated by sampling around a theoretically computed data vector with our simulated covariance matrix. We compute the p-value, defined as the probability that the $\chi^2$ from a random realization will exceed the value obtained with Y3 data. We find p=0.14. The distribution of $\chi^2$ values is shown along with the Y3 value in Figure \ref{fig:chi2dist_unblind}. The average $\chi^2$ for a noisy mock simulation is 73.3. While this value is different than the number of degrees of freedom computed through the method by \citet{Raveri.Hu.2019}, both approaches show the statistical validity of our joint Y3 result. The differences between the results arise from the fact that the degrees-of-freedom approach assumes a Gaussian approximation for the posterior. The mock approach is free of this assumption, but its precision can be affected by the number of realizations.

We also visually confirm that our best fit matches well our $\xi_{\pm}$ and $\mapcube$ data points at all redshifts and scales, without any systematic departure from the fit. We show this result in Figures \ref{fig:meas} (for $\mapcube$) and \ref{fig:meas_2pcf} (for $\xi_{\pm}$).

We also use our best fit cosmology to predict the $\xi_{\pm}$ and $\mapcube$ at scales excluded from the fit. This prediction is made with a dark-matter-only theoretical model, in order to verify if significant departures from the data would be identifiable at small scales. However, we still find adequate concordance both in $\mapcube$ (Figure \ref{fig:meas2}) and $\xi_{\pm}$ (Appendix~\ref{sec:appendix-ss2pcf}). This serves as an independent validation of the robustness of the joint $\xi_{\pm} + \mapcube$ analysis against baryonic effects. In our previous methodology paper, we show that this robustness is achieved by selecting scales on $\mapcube$ in which the net effect of baryons is one of a slight enhancement of the signal. The combination of these scales with the known suppression on the 2PCF makes the joint analysis more robust against shifts on $S_8$.

While in our methodology paper we had forecast a slight improvement on the intrinsic alignment amplitude parameter, we do not see it in Y3 data. The question remains whether the future development and validation of more complex IA modeling for three-point statistics will allow us to find a pattern of improvement when moving from $\xi_{\pm}$ to $\xi_{\pm}$+$\mapcube$.

\subsection{$w$CDM Cosmological Constraints}\label{sec:result-des-y3-wcdm}

We also perform a run of our pipeline with $w$CDM modeling, and find an improvement of 66\% on the $\Omega_m$-$S_8$ FoM. Our improvement in the joint $S_8$-$w_0$ constraint is of 22\%. This is higher than the 5\% improvement we see on the marginalized $w_0$ parameter. This difference is compatible with what we identified in our methodology paper, that the inclusion of mass aperture allows a degeneracy breaking in the $S_8$-$w_0$ plane. We show our results in Figure \ref{fig:y3wcdm}, with the mean posteriors listed on Table~\ref{tab:param_values}. When including $\mapcube$ in the $w$CDM analysis, our $w_0$ constraint remains compatible with $w_0=-1$ at approximately $1\sigma$.

While our analysis finds a better improvement on $\Omega_m$-$S_8$ in $\Lambda$CDM rather than $w$CDM, this is not a pattern across all Y3 HOS analyses. The simulation-based inference (SBI) by \citet{Gatti.Campailla.2024} reports a gain of $70\%$ with $\Lambda$CDM and $90\%$ with $w$CDM when adding third moments, scattering transforms and wavelet phase harmonics to the second moment of the convergence field. Another SBI result by \citet{Jeffrey.Whiteway.Gatti.2024} combines power spectra, peak counts, and a direct map-level inference, finding for $w$CDM improvements of $126\%$ on the $\Omega_m$-$S_8$ FoM and $148\%$ on the $\Omega_m$-$w_0$ FoM. A main difference between these analyses and the present work is that both of them require $-1<w_0<-1/3$, while we use wider priors and allow for $w_0 < -1$. Since most of our posterior distribution is situated on the $w_0 < -1$ regime, we are essentially probing a distinct part of the parameter space.

In order to directly compare the DES Y3 $\xi_{\pm}$+$\mapcube$ $w$CDM cosmological constraints with those from other higher-order statistics that require simulation-based inference, we perform an additional run of our pipeline with adapted priors to match those of \citet{Gatti.Campailla.2024}, \citet{Jeffrey.Whiteway.Gatti.2024}, and \citet{Prat.Gatti.Doux.2025}. In this analysis, we restrict the dark energy equation of state parameter to $-1<w_0<-1/3$, and introduce Gaussian priors on $h_0$, $n_s$, and $\Omega_bh_0^2$, as described in Table I of \citet{Jeffrey.Whiteway.Gatti.2024}. We present, in Figure \ref{fig:HOS}, the DES Y3 $\xi_{\pm}$+$\mapcube$ constraints on $\Omega_m$, $S_8$, and $w_0$, together with those from third moments, scattering transforms, and wavelet phase harmonics \cite{Gatti.Campailla.2024}, persistent homology (Betti numbers) \cite{Prat.Gatti.Doux.2025}, and power spectrum, peaks, and map-level inference with CNNs \cite{Jeffrey.Whiteway.Gatti.2024}. We also include as a baseline the $\xi_{\pm}$ result from \citet{Prat.Gatti.Doux.2025}, which adopts the same priors as those from the SBI analyses. All the constraints from these HOSs are consistent with each other and represent, individually, significant improvements relative to $\xi_{\pm}$ alone.

\subsection{Comparison with external CMB data}

We compare our results with CMB constraints from the Planck Collaboration in order to investigate whether they affect the current tension between early and late universe probes of $S_8$. Our results are shown in Figure \ref{fig:s8vsplanck}. We quantify the tension in the $S_8$-$\Omega_m$ plane through the method devised by \citet{Raveri.Hu.2019}. We obtain, for the tension between DES Y3 $\xi_{\pm}$ and Planck 2018, a significance of $2.3\sigma$. The tension between DES Y3 $\mapcube$ alone and Planck is of $1.9\sigma$. Finally, the tension between our joint constraint and the one from Planck is of $2.3\sigma$, remaining the same as when we use the two-point only data set. 

\subsection{Uninformative priors on photo-z shift parameter}
\label{sec:photoz}

\begin{figure*}
    \centering
    \includegraphics[width=\linewidth]{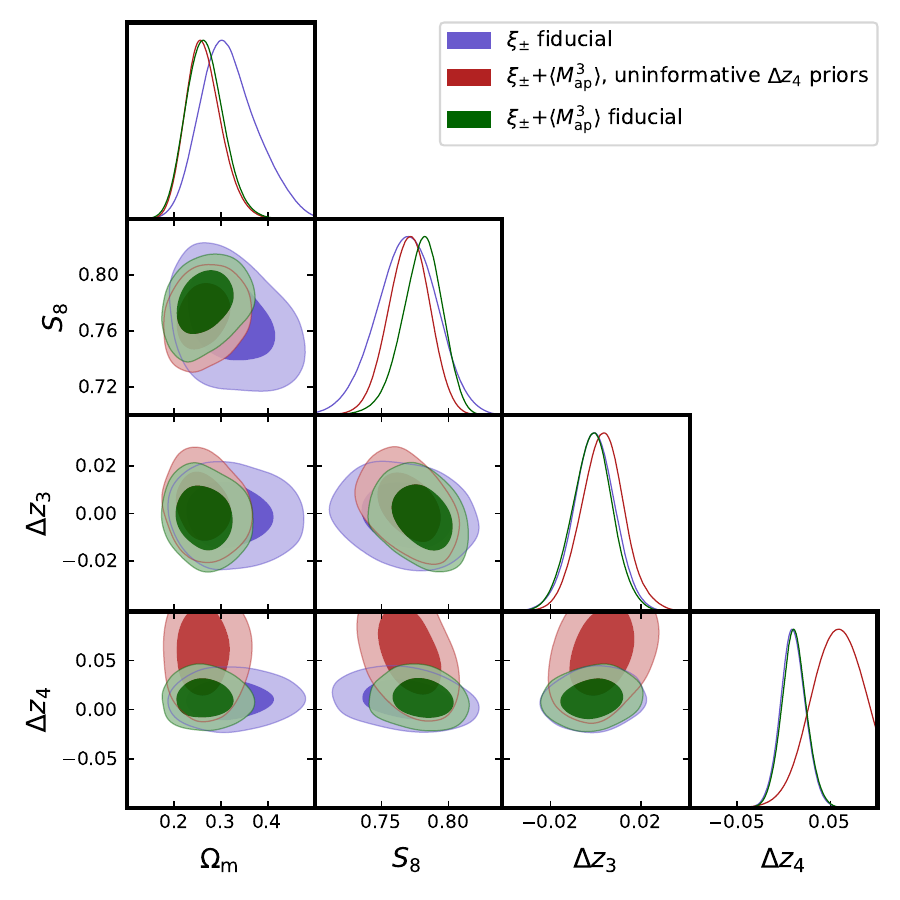}
    \caption{Parameter constraints from $\xi_{\pm}$ and $\mapcube$ analysis using uninformative flat priors on the photo-z shift parameter $\Delta z_4$. We find that by removing the informative prior, the joint $\xi_{\pm}$-$\mapcube$ analysis calibrates $\Delta z_4$ to a value higher than the one assumed in the fiducial DES analyses, leading to a shift towards lower $S_8$. We also include the third photo-z shift parameter $\Delta z_3$ to indicate the slight degeneracy between it and $\Delta z_4$.}
    \label{fig:flatdz34priors}
\end{figure*}

Our main cosmological analysis follows \cite{Secco.To.2021} and \cite{Amon.Weller.2021} in using informative priors on the photo-z shift parameters. Here we also run our pipeline removing the informative prior on $\Delta z_4$ and replacing it with a flat prior ranging from -0.3 to 0.3. This is motivated by the observation that the density of galaxies used for photo-z calibration of the fourth z-bin is small, which can lead to larger potential biases. We find that by using an uninformative prior on $\Delta z_4$, our pipeline finds its value to be significantly higher than that of the fiducial Y3 analysis, being 2$\sigma$ above zero (Figure~\ref{fig:flatdz34priors}). Due to the degeneracy of the photo-z shifts with $S_8$, we find a preference for a lower $S_8$ relative to the joint fiducial result. We obtain $S_8=0.770\pm 0.015$. While the $\xi_{\pm}$ result alone is degraded by removing the informative priors, the contours do not get larger for the joint analysis, which yields us a FoM improvement of $144\%$ on the $\Omega_m - S_8$ plane. As a result of the shift in $S_8$, the tension between the joint chain with uninformative $\Delta z_4$ priors and CMB results from Planck 2018 is slightly larger, with a significance of $2.6\sigma$.

We also perform the same analysis procedure on a simulated data vector and obtain for the mean photo-z shifts results consistent (at 0.4$\sigma$) with their true values. Therefore, the observed shift on Y3 data should not be dismissed as resulting from projection effects alone.

\section{Conclusion}\label{sec:conclusion}

We measure the full configuration-dependent three-point correlation function on the DES Y3 shape catalog. We perform the first set of tomographic three-point measurements over all 20 combinations of the four fiducial source galaxy redshift bins. We  compress these measurements into the mass aperture skewness statistic with essentially no loss of information. We also perform new measurements of the two-point correlation function on the shape catalog, correcting for an issue on the data vector of the DES fiducial analysis. We extend our measurements to small scales (down to $0.25'$) for validation purposes (see Appendix~\ref{sec:appendix-ss2pcf} for a test of baryonic feedback effects).

We perform a $\Lambda$CDM cosmological analysis with two-point and three-point correlation functions measured on the DES Y3 shape catalog, through a joint $\xi_{\pm}$ - $\mapcube$ data vector. We find $S_8=0.780\pm0.015$ and $\Omega_{\rm m}=0.266^{+0.039}_{-0.040}$. We obtain an improvement of $111\%$ on the figure of merit on the $\Omega_m$-$S_8$ plane. This level of improvement is driven by the slightly different degeneracy directions of the individual 2PCF and $\mapcube$ contours, as can be seen in Figure \ref{fig:y3result_full}. This significant increase in cosmological information is somewhat surprising in view of other higher order statistics applied to data (\cite{Gatti.Wiseman.2023}), but is consistent with that found by the KiDS collaboration \cite{Burger.Martinet.2023} and by our simulated analysis. Much of the gain comes from using the full configuration dependence of the three-point function. We also test the $w$CDM model and find a more modest $22\%$ improvement in the joint $S_8$ - $w_0$ figure of merit. Our constraints from $w$CDM are $\Omega_{\rm m}=0.242^{+0.038}_{-0.037}$, $S_8=0.749^{+0.027}_{-0.026}$ and $w_0=-1.39\pm 0.31$. 

We compare our results with primary CMB data and find that the tension between previous LSS results from DES and CMB results from Planck 2018 persists with addition of the new data set. We find a tension of $2.3\sigma$ between our joint $\xi_{\pm}$+$\mapcube$ results and Planck 2018. While our constraints were significantly tightened in parameter space, the persistence of the tension motivates further work towards DES Year 6 data and methods to use all the available information in the shear field.

We have found that the addition of $\mapcube$ makes the analysis more robust to baryonic feedback 
\cite[See Fig.~10 of][]{gomes2025cosmologysecondthirdordershear}. 
This finding needs further study with a wider set of hydrodynamic simulations and variation in scale cuts. It opens the possibility of designing a joint analysis that improves constrains on both cosmological and baryonic feedback parameters. 

We explored variations in the treatment of uncertainty due to photo-z estimation. Section~\ref{sec:photoz} shows the results of freeing up the prior on the mean redshift of the fourth bin, which has limited coverage in the calibrating spectroscopic redshifts. We find that the joint analysis can constrain the mean, with the posterior showing a  shift of about 0.05 towards higher redshift. This results in a slightly lower $S_8$ and increases the tension with Planck to 2.6$\sigma$. We also tested the effect of freeing up both the third and fourth redshift bin, with similar conclusions. 

The assessment of the tension with Planck can also potentially be affected by the choice of non-linear power spectrum prescription. \citet{anbajagane2025decadecosmicshearproject} report a small shift towards higher $S_8$ (and lower tension) when using \software{HMCode} \cite{HMCode2020} instead of Halofit. A similar shift was noticed by \citet{Secco.To.2021}, who attributed it to a projection effect of their free baryon parameters. We leave this topic to be explored in further study.

The methodology for marginalizing over systematics can be improved in future work. In particular, for modeling IA, we have restricted this study to the NLA model. While there are suggestions from previous DES, KiDS and HSC studies that the amplitude of this signal is modest (or consistent with zero), the TATT model is a more realistic model and also complex -- it would lead to weaker cosmological constraints. Simulation-based models of IA are also emerging and would provide a useful cross-check \cite{VanAlfen:2023rew}. At the level of precision of DES Y3, other systematics such as reduced shear or magnification are expected to be negligible, but these will need a more careful examination for Stage 4 surveys. 

\section{Author Contributions}

RCHG and SS developed the inference pipeline, performed the cosmological analyses, and composed the manuscript. BJ served as project advisor and contributed to the manuscript. MJ developed novel \software{TREECORR} functionalities that were used in this work. DA generated the \software{CosmogridV1} mocks used in this work. AH, GAM and SP served as collaboration internal reviewers. JM was the final publication reader. MG, ES, AAm, MB, MT, ACh, CDo, NM, ANA, IH, DG, GB, MJ, LFS, AF, TS, JMc, RPR, RCh, CC, SP, IT, JP, JEP, CS
are credited for the development of the DES shape catalog; JMy, AAl, AAm, CS, SE, JD, JMc, DG, GB, MT, SD, ACa, NM, BYi, MR, MG, GG, RCa, AJR, ESR, JEP, JC, IH, JP
for the production of the DES source redshift distributions; SE, BYa, NK, EMH, YZ for the development of the DES Balrog; and NM, MB, JMc, AA, DG, MJ, ACh, MT, ES, BYa, KH, SD, JZ, KE, RPR, TNV for the development of the DES image simulations. JMu, GB, DH, FE are credited for the development of the catalog blinding; MJ, GB, AA, CDa, PFL, KB, IH, MG, AR for the DES PSF;
and NM and JZ for the development of the \software{CosmoSIS} framework.

\section{Acknowledgments}

BJ and RCHG are  partially supported by the US Department of Energy grant DE-SC0007901 and SS is supported by the JSPS Overseas Research Fellowships. Part of this work was supported by the NASA ROSES grant 22-ROMAN11-0011 via a JPL subaward.

Funding for the DES Projects has been provided by the U.S. Department of Energy, the U.S. National Science Foundation, the Ministry of Science and Education of Spain, 
the Science and Technology Facilities Council of the United Kingdom, the Higher Education Funding Council for England, the National Center for Supercomputing 
Applications at the University of Illinois at Urbana-Champaign, the Kavli Institute of Cosmological Physics at the University of Chicago, 
the Center for Cosmology and Astro-Particle Physics at the Ohio State University,
the Mitchell Institute for Fundamental Physics and Astronomy at Texas A\&M University, Financiadora de Estudos e Projetos, 
Funda{\c c}{\~a}o Carlos Chagas Filho de Amparo {\`a} Pesquisa do Estado do Rio de Janeiro, Conselho Nacional de Desenvolvimento Cient{\'i}fico e Tecnol{\'o}gico and 
the Minist{\'e}rio da Ci{\^e}ncia, Tecnologia e Inova{\c c}{\~a}o, the Deutsche Forschungsgemeinschaft and the Collaborating Institutions in the Dark Energy Survey. 

The Collaborating Institutions are Argonne National Laboratory, the University of California at Santa Cruz, the University of Cambridge, Centro de Investigaciones Energ{\'e}ticas, 
Medioambientales y Tecnol{\'o}gicas-Madrid, the University of Chicago, University College London, the DES-Brazil Consortium, the University of Edinburgh, 
the Eidgen{\"o}ssische Technische Hochschule (ETH) Z{\"u}rich, 
Fermi National Accelerator Laboratory, the University of Illinois at Urbana-Champaign, the Institut de Ci{\`e}ncies de l'Espai (IEEC/CSIC), 
the Institut de F{\'i}sica d'Altes Energies, Lawrence Berkeley National Laboratory, the Ludwig-Maximilians Universit{\"a}t M{\"u}nchen and the associated Excellence Cluster Universe, 
the University of Michigan, NSF NOIRLab, the University of Nottingham, The Ohio State University, the University of Pennsylvania, the University of Portsmouth, 
SLAC National Accelerator Laboratory, Stanford University, the University of Sussex, Texas A\&M University, and the OzDES Membership Consortium.

Based in part on observations at NSF Cerro Tololo Inter-American Observatory at NSF NOIRLab (NOIRLab Prop. ID 2012B-0001; PI: J. Frieman), which is managed by the Association of Universities for Research in Astronomy (AURA) under a cooperative agreement with the National Science Foundation.

The DES data management system is supported by the National Science Foundation under Grant Numbers AST-1138766 and AST-1536171.
The DES participants from Spanish institutions are partially supported by MICINN under grants PID2021-123012, PID2021-128989 PID2022-141079, SEV-2016-0588, CEX2020-001058-M and CEX2020-001007-S, some of which include ERDF funds from the European Union. IFAE is partially funded by the CERCA program of the Generalitat de Catalunya.

We  acknowledge support from the Brazilian Instituto Nacional de Ci\^encia
e Tecnologia (INCT) do e-Universo (CNPq grant 465376/2014-2).

This document was prepared by the DES Collaboration using the resources of the Fermi National Accelerator Laboratory (Fermilab), a U.S. Department of Energy, Office of Science, Office of High Energy Physics HEP User Facility. Fermilab is managed by Fermi Forward Discovery Group, LLC, acting under Contract No. 89243024CSC000002.

This research used resources of the National Energy Research Scientific Computing Center (NERSC), a Department of Energy User Facility using NERSC award DOE-ERCAP0031464

\appendix

 \section{Comparison with other DES two-point analyses}
For this work, we performed our own measurement and analysis of the two-point correlation function on DES Y3 data. Here we show the differences in the analysis choices between our pipeline and the fiducial Y3 pipeline. 

Firstly, \citet{mccullough2024darkenergysurveyyear} identified a mismatch between the tomographic binning file used in the Y3 fiducial analysis and an outdated version  which was present on the shear catalog file and used to produce the two-point data vector. Following their procedure, we created a new two-point data vector making use of the same shear catalog but updating the tomographic binning to match the redshift distributions assumed by the pipeline (SOMPZ version 0.5). 

\begin{figure}
    \centering
    \includegraphics[width=\linewidth]{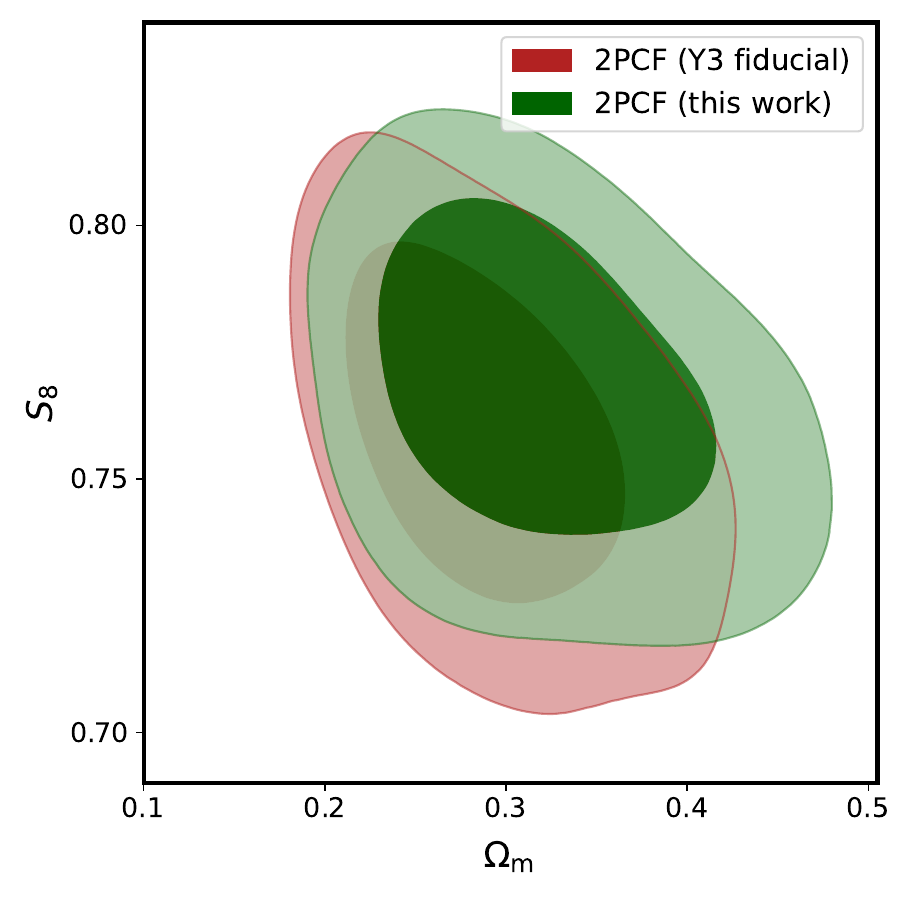}
    \caption{Results from our two-point shear analysis along with the fiducial Y3 cosmic shear results. The differences between our chains include our correction of the tomographic binning as well as methodological differences such as the use of NLA for intrinsic alignments. The shift in $S_8$ between both analyses is of 0.2$\sigma$.}
    \label{fig:fiducial_vs_ours}
\end{figure}

Next, our analysis also differs from the Y3 fiducial contours in that we use NLA instead of TATT for intrinsic alignment modeling, in order to have a consistent modeling between $\xi_{\pm}$ and $\mapcube$. This is what drives most of the difference between the contours. We do not use any additional shear ratio data vector, and we adapt our parameter space sampling to match with our emulator support range as described in Paper~I. Finally, we make use of the Percival likelihood to better propagate the uncertainties of our simulated covariance into the final constraints (Eq \ref{eq:percival}). We show our contours alongside the DES Y3 fiducial ones on Figure \ref{fig:fiducial_vs_ours}, obtaining a small shift of 0.2$\sigma$ on the mean $S_8$ value.

\section{Blind analysis}
\label{sec:appendix-blind}

Previous to running our chains on the actual Y3 data vector, we performed a blind analysis, by both making use of a shifted catalog, such as was done in previous Y3 analyses, and by hiding the parameter values on our plots and showing only shifts relative to the $\xi_{\pm}$ mean posterior values. The only methodological change made after unblinding was the shift from a Gaussian likelihood to the one proposed by \citet{Percival.Friedrich.2021}. While this affects the shape of the posterior tail, our blind tests remain valid due to the use of posterior peaks and confidence intervals.

We generated 50 noisy realizations at CosmoGridV1 cosmology, using our measured covariance matrix to generate Gaussian noise. We ran our pipeline on each of the simulations in order to verify the distributions of $\chi^2$, of $\Omega_m$-$S_8$ figure-of-merit, and of the shift in cosmological parameters when moving from a two-point only analysis to a joint analysis. Our $\chi^2$ distribution is the same shown in Figure \ref{fig:chi2dist_unblind}. The blind $\chi^2$ value is 69.7, which gives a p-value of 0.58. 

\begin{figure}
    \centering
    \includegraphics[width=\linewidth]{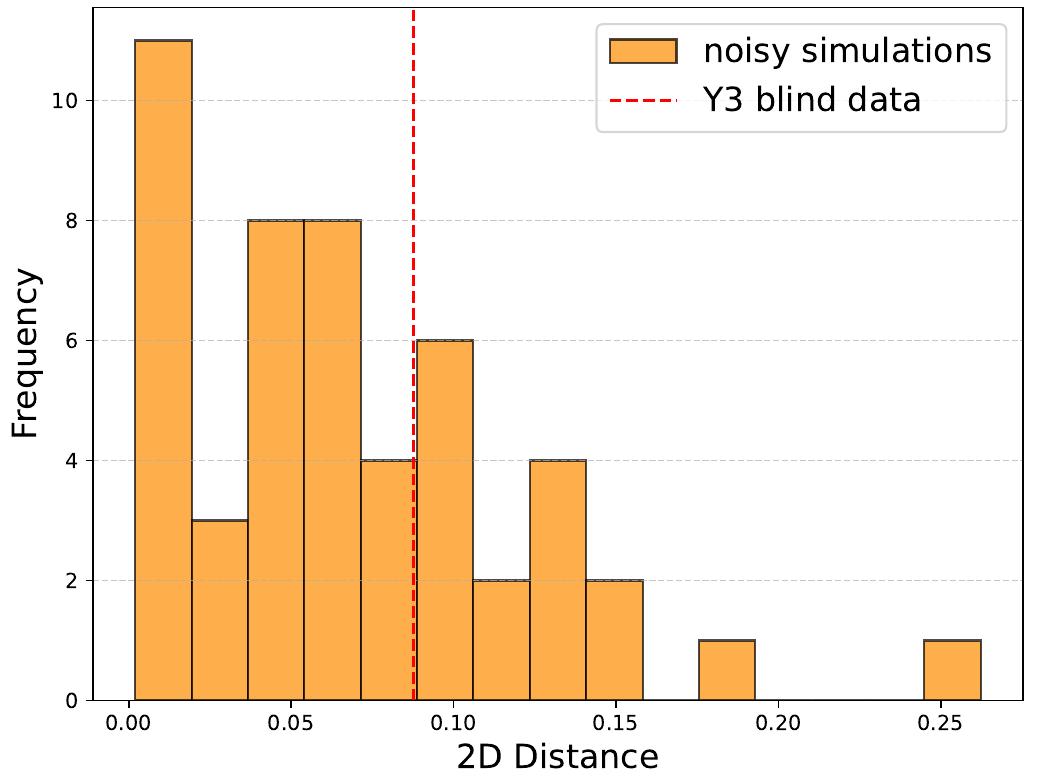}
    \caption{Two-dimensional distance between the $\Omega_m$-$S_8$ posterior means when shifting from two-point only to a joint analysis. The histogram shows the results for 50 different noise realizations, while the red dashed line indicates the shift obtained in blind Y3 data. We verify that the Y3 data shift falls within the expected distribution, with a p-value of 0.32. After unblinding, the result remained consistent, with a p-value of 0.78.}
    \label{fig:shiftparams}
\end{figure}

We also compute, for each one of our noisy simulations, the 2D parameter space distance between the peak of the $\Omega_m$-$S_8$ posteriors from the two-point only and from the joint analysis. We show the distribution along with the Y3 inferred value in Figure \ref{fig:shiftparams}. The shift found in data also lies within the expected distribution, with a p-value of 0.32.

Finally, we verify the figure of merit on the $\Omega_m$-$S_8$
 plane for each of the noisy simulations in both the $\xi_{\pm}$ scenario and the $\xi_{\pm}$ + $\mapcube$. The probabilities of exceeding are 88\% for $\xi_{\pm}$ and 40\% for the joint chain. When we analyze the distribution of the figure-of-merit ratios, we find an improvement range of between 24\% to 170\%. This indicates that the factor-of-two increase in constraining power when adding $\mapcube$  to $\xi_{\pm}$ can be significantly suppressed or boosted depending on the noise realization.

\section{Full set of contours for the cosmological parameters}
\label{sec:allcontours}
\begin{figure*}
    \centering
    \includegraphics[width=\linewidth]{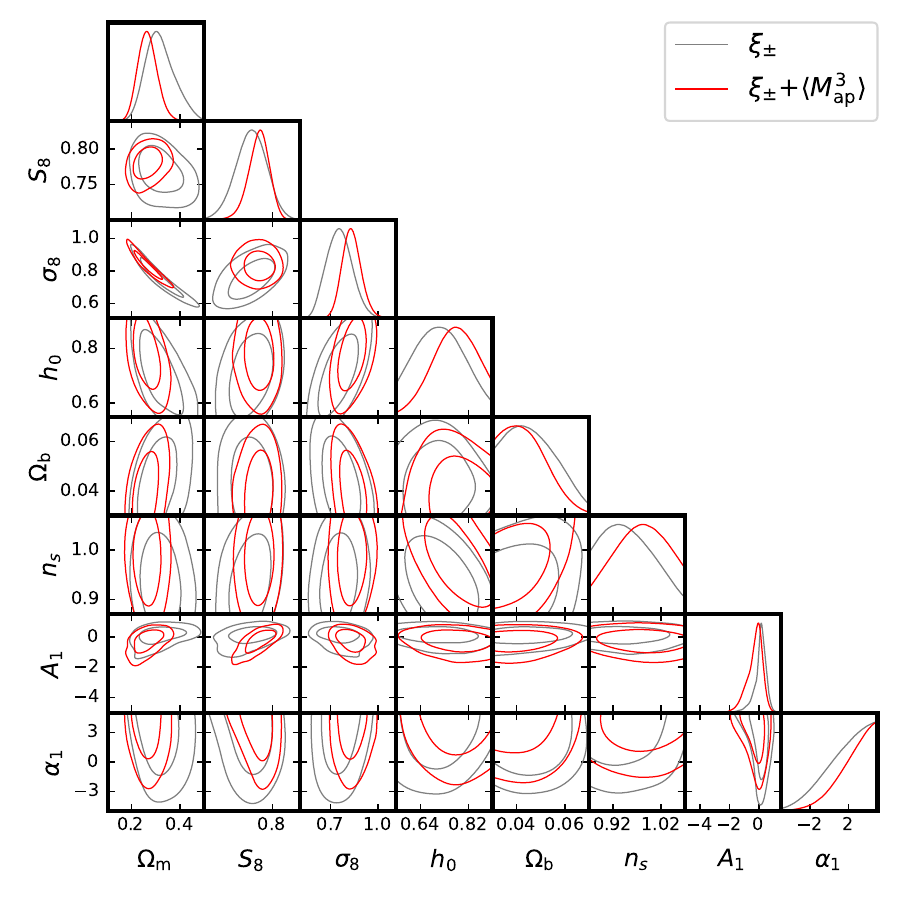}
    \caption{Parameter constraints on the full set of cosmological parameters from $\xi_{\pm}$ and $\mapcube$ using $\Lambda$CDM modeling. The dark blue contours indicate results from the 2PCF alone, while the red contours indicate results from the 2PCF combined with $\mapcube$.}
    \label{fig:lcdm_allparams}
\end{figure*}

\begin{figure*}
    \centering
    \includegraphics[width=\linewidth]{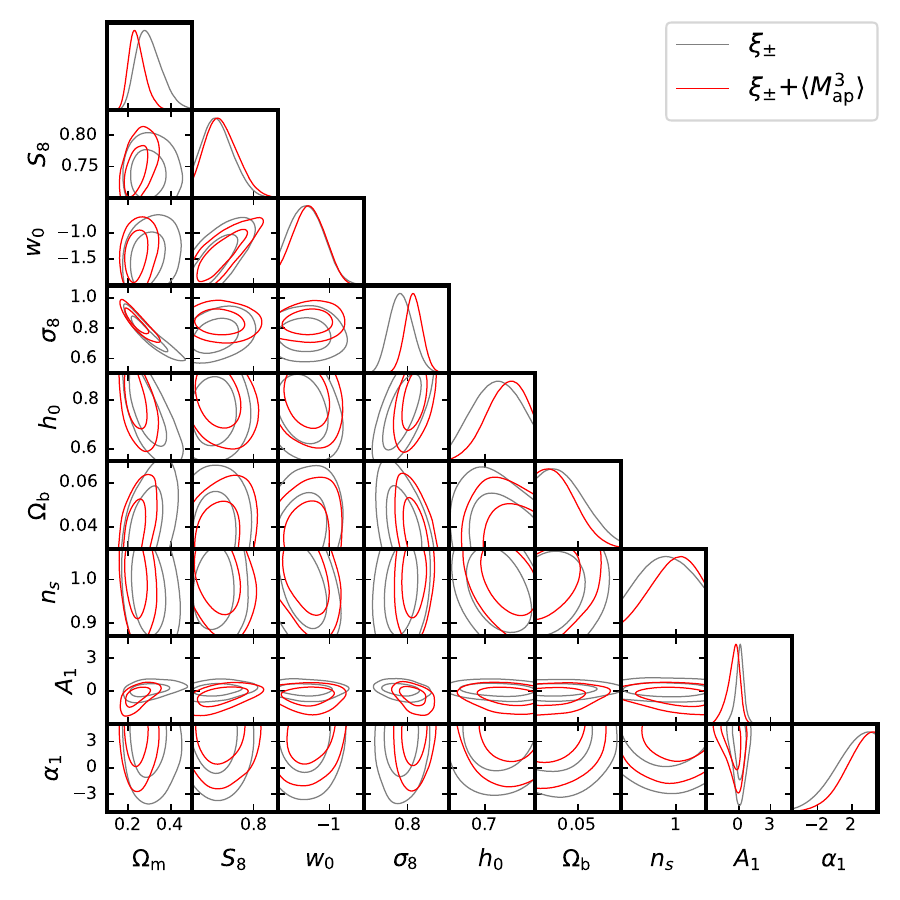}
    \caption{Parameter constraints on the whole set of cosmological parameters from $\xi_{\pm}$ and $\mapcube$ using $w$CDM modeling. The dark blue contours indicate results from the 2PCF alone, while the red contours indicate results from the 2PCF combined with the mass aperture statistic.}
    \label{fig:wcdm_allparams}
\end{figure*}

Here we present our full set of contours obtained with both $\Lambda$CDM and $w$CDM modeling. We do not include constraints on the sum of the neutrino masses because our $\mapcube$ emulator assumes a fixed value of $\Sigma m_{\nu}=0.06$. For $\Lambda$CDM, our results are shown in Figure~\ref{fig:lcdm_allparams}. For $w$CDM, we present our results in Figure~\ref{fig:wcdm_allparams}.

\section{Small scale measurements of the 2PCF}

\label{sec:appendix-ss2pcf}
Our cosmological analysis makes use of the same 2PCF measurements as done in the fiducial Y3 analyses, spanning from $2.5'$ to $250'$ in 20 log-spaced bins. Some of the smallest scales are removed to avoid biasing due to baryonic suppression. 

\begin{figure*}
    \centering
    \includegraphics[width=\linewidth]{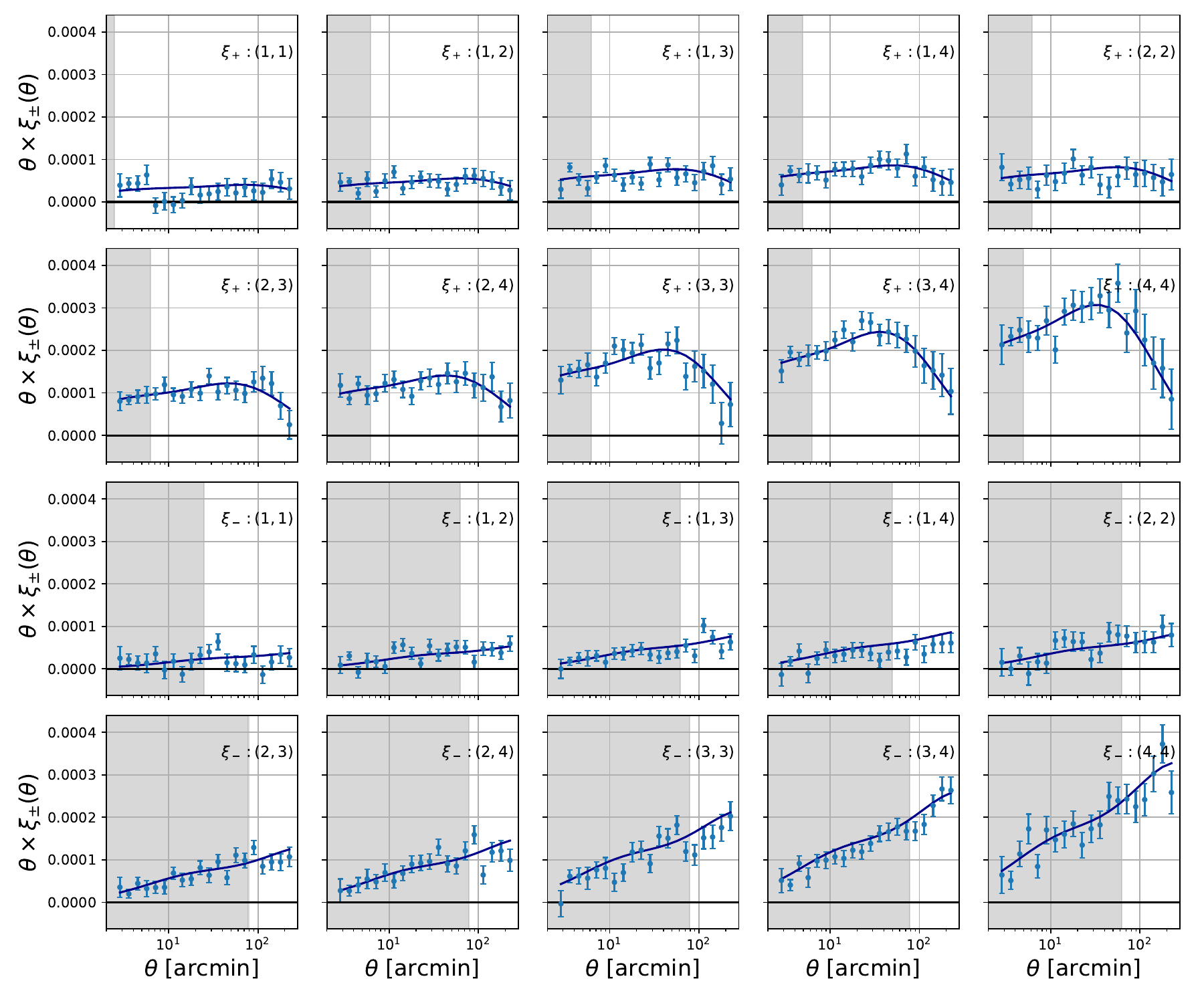}
    \caption{2PCF Y3 data along with the theoretical prediction at our joint best fit cosmology. The gray region corresponds to data points which were not used in the fit due to scale cuts. We verify that the joint dark matter-only theory model reasonably fits the excluded data points. A more detailed analysis is required in order to identify possible deviations in the data due to baryonic feedback and to quantify its level of agreement with the dark matter-only model.}
    \label{fig:meas_2pcf}
\end{figure*}

\begin{figure*}
    \centering
    \includegraphics[width=\linewidth]{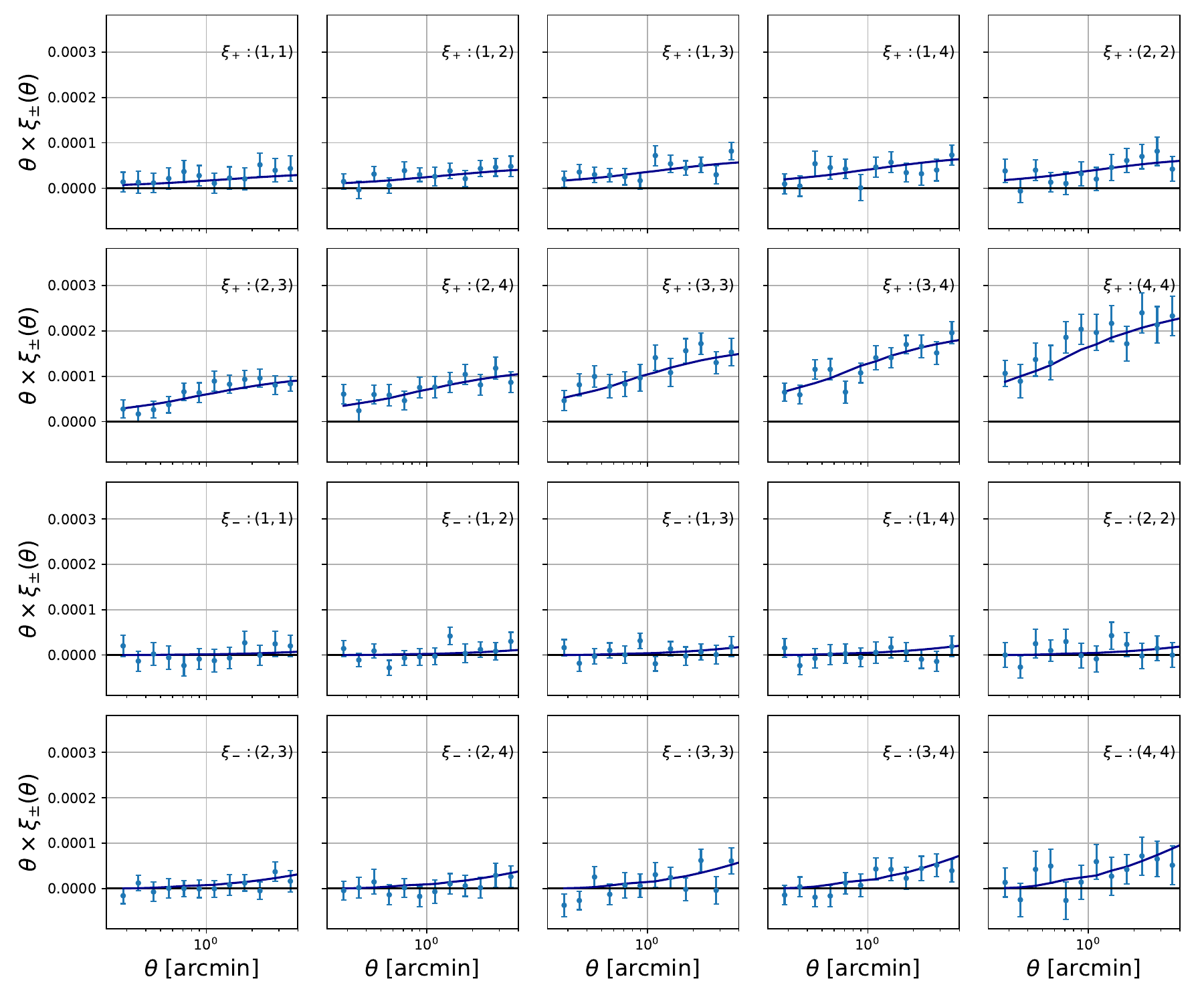}
    \caption{2PCF Y3 data at scales around 1 arcmin along with the theoretical dark matter-only prediction at joint best fit cosmology (computed with fiducial Y3 scale cuts). At our level of noise, the baryonic supression of the signal is not clearly identifiable.}
    \label{fig:meas_2pcf_small}
\end{figure*}

We perform a new set of measurements at even smaller scales, not to be used in our fit, but to verify whether a significant deviation from dark-matter-only theoretical predictions would be apparent from Y3 shear data alone. We introduce 10 new log-spaced bins between $0.25'$ and $2.5'$, and compute their error bars via jackknife sampling of the data. The fiducial Y3 measurements and our new small scale measurements are shown in Figures \ref{fig:meas_2pcf} and \ref{fig:meas_2pcf_small} along with a Halofit prediction using our joint best fit parameters. We find the joint cosmology to be still a reasonable fit to the small scale data. 

A similar analysis by \citet{pandey2025} found the Y3 2PCF small scale data points to be significantly below the Halofit theoretical prediction. This result, however, differs from ours in that their best fit parameters are obtained from a joint DES Y3 and ACT DR6 tSZ analysis with baryonic free parameters, which yields an $S_8$ value much higher than ours.

\bibliography{refs}

\end{document}